\documentclass[sigconf]{acmart}
\AtBeginDocument{%
  }

\def\urlicon#1{\href{#1}{\includegraphics[width=7pt]{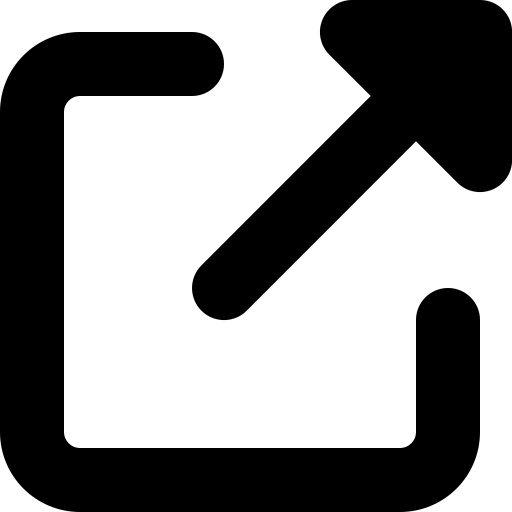}}}%

\usepackage{booktabs}
\usepackage{multirow}
\usepackage{algorithm}
\usepackage{algpseudocode}
\usepackage{circledsteps}
\usepackage{tabularx}
\usepackage{enumitem}
\usepackage{cleveref}
\usepackage{wrapfig}
\usepackage{tablefootnote}
\usepackage{footnote}
\makesavenoteenv{tabular}
\makesavenoteenv{table}

\usepackage[english]{babel}

\algnewcommand{\IfThenElse}[3]{% \IfThenElse{<if>}{<then>}{<else>}
  \State \algorithmicif\ #1\ \algorithmicthen\ #2\ \algorithmicelse\ #3}
  
\copyrightyear{2025}
\acmYear{2025}
\setcopyright{acmlicensed}\acmConference[ISCA '25]{Proceedings of the 52nd Annual International Symposium on Computer Architecture}{June 21--25, 2025}{Tokyo, Japan}
\acmBooktitle{Proceedings of the 52nd Annual International Symposium on Computer Architecture (ISCA '25), June 21--25, 2025, Tokyo, Japan}
\acmDOI{10.1145/3695053.3731021}
\acmISBN{979-8-4007-1261-6/2025/06}
\begin{document}
% \input{-01-cover-letter}
\title[Need for zkSpeed: Accelerating HyperPlonk for Zero-Knowledge Proofs]{Need for zkSpeed: \\ 
Accelerating HyperPlonk for Zero-Knowledge Proofs 
}
\author{Alhad Daftardar}
\affiliation{
    \institution{New York University \\ Tandon School of Engineering}
    \city{Brooklyn}
    \state{NY}
    \country{USA}
}
\email{ajd9396@nyu.edu}

\author{Jianqiao Mo}
\affiliation{
    \institution{New York University \\ Tandon School of Engineering}
    \city{Brooklyn}
    \state{NY}
    \country{USA}
}
\email{jm8782@nyu.edu}

\author{Joey Ah-kiow}
\affiliation{
    \institution{New York University \\ Tandon School of Engineering}
    \city{Brooklyn}
    \state{NY}
    \country{USA}
}
\email{ja4844@nyu.edu}

\author{Benedikt B{\"u}nz}
\affiliation{
    \institution{New York University \\ Courant Institute}
    % \city{Manhattan}
    \city{New York}
    \state{NY}
    \country{USA}
}
\email{bb@nyu.edu}

\author{Ramesh Karri}
\affiliation{
    \institution{New York University \\ Tandon School of Engineering}
    \city{Brooklyn}
    \state{NY}
    \country{USA}
}
\email{rkarri@nyu.edu}

\author{Siddharth Garg}
\affiliation{
    \institution{New York University \\ Tandon School of Engineering}
    \city{Brooklyn}
    \state{NY}
    \country{USA}
}
\email{sg175@nyu.edu}

\author{Brandon Reagen}
\affiliation{
    \institution{New York University \\ Tandon School of Engineering}
    \city{Brooklyn}
    \state{NY}
    \country{USA}
}
\email{bjr5@nyu.edu}


% \affiliation{%



\begin{abstract}

Zero-Knowledge Proofs (ZKPs) are a rapidly growing technique for privacy-preserving and verifiable computation.
ZKPs enable one party (a prover: $\mathcal{P}$) to prove to another (a verifier: $\mathcal{V}$) that a statement is true or correct without revealing any additional information.
This powerful capability has led to ZKPs being applied and proposed for application in blockchain technologies, verifiable machine learning, and electronic voting.
However, ZKPs have yet to see widespread, ubiquitous adoption due to the exceptionally high computational complexity of the proving process.
Naturally, there has been recent work to accelerate ZKP primitives and protocols using GPUs and ASICs.
However, the protocols considered so far face one of two challenges:
they require a trusted setup for each new application or
generate large proofs with high verification costs, limiting their applicability in scenarios with numerous verifiers or strict verification time constraints.
HyperPlonk is a state-of-the-art ZKP protocol that supports both one-time, universal setup and small proof sizes/verification costs expected by publicly verifiable, consensus-based systems (e.g., blockchain).
While HyperPlonk's setup and verifier properties are highly desirable, the proving phase is costly.
A HyperPlonk prover must compute on large bitwidths (e.g., 255-381b) and polynomials (e.g., of degree 2$^{24}$),
employs computationally (e.g., MSM) and bandwidth (e.g., SumCheck) intensive kernels,
and the complete protocol comprises many steps, each constituting distinct kernels.
We present an accelerator, \emph{zkSpeed}, to address these challenges and effectively accelerate HyperPlonk.
zkSpeed provides hardware support for all major primitives (e.g., SumCheck and Multi-Scalar Multiplications (MSMs)) and judiciously schedules each protocol phase onto the allocated hardware. 
We leverage high-level synthesis to thoroughly explore and optimize the hardware design tradeoffs of each unit.
These are then input into a full-chip simulator for large-scale design space exploration to optimize all aspects of the architecture in unison.
Our Pareto analysis demonstrates that with a 366mm$^2$ chip and 2 TB/s of off-chip bandwidth, zkSpeed is able to accelerate the entire proof generation by 801$\times$ (geomean) over a CPU baseline.

\end{abstract}

\begin{CCSXML}
<ccs2012>
   <concept>
       <concept_id>10010520.10010521.10010542.10011714</concept_id>
       <concept_desc>Computer systems organization~Special purpose systems</concept_desc>
       <concept_significance>500</concept_significance>
       </concept>
   <concept>
       <concept_id>10002978.10002979</concept_id>
       <concept_desc>Security and privacy~Cryptography</concept_desc>
       <concept_significance>500</concept_significance>
       </concept>
 </ccs2012>
\end{CCSXML}

\ccsdesc[500]{Computer systems organization~Special purpose systems}
\ccsdesc[500]{Security and privacy~Cryptography}
% \keywords{datasets, neural networks, gaze detection, text tagging}
\keywords{Zero-Knowledge Proofs, Cryptography, Hardware Acceleration}

\maketitle

\vspace{-.5em}
\section{Introduction}\label{sec:intro}

Zero-Knowledge Proofs (ZKPs) are a privacy-preserving computation technique that enable one party (a prover $\mathcal{P}$)
to prove to another (a verifier $\mathcal{V}$) that a computation (possibly with secret inputs) was performed correctly. 
The proof itself reveals no information about the (secret) inputs nor the function, 
and it can be verified much faster than the compute time of the original computation.
Proof systems can be characterized by several important criteria, including: 
prover time, verifier time, proof size, protocol setup, and cryptographic assumptions required for the proof to be secure.
Each deployment of ZKPs can result in a different set of criteria requirements. 
For example, private transactions~\cite{zerocash} require one proof per transaction be posted on a blockchain and distributed to all blockchain nodes, thus prioritizing small proof size.

Recent years have seen significant advances and heterogeneity in proof systems~\cite{groth,plonk,hyperplonk,orion,spartan,10.1145/3658644.3670316}. 
Each optimizes for a distinct set of criteria constraints, and we note that in all systems, prover time is vital to optimize for,
as faster provers enable new applications and proving is much slower than verification~\cite{libsnark, hyperplonk}.
To understand the ZKP space, consider Groth16~\cite{groth} and Orion~\cite{orion}, 
two ZKPs with distinct characteristics that have also recently been accelerated~\cite{pipezk, szkp, nocap}.
Groth16 produces very short proofs (192 bytes) and enables millisecond verification latency, independent of proved computation size. 
However, it relies on a strong cryptographic assumption: a \emph{circuit-specific trusted setup} where a trusted party generates keys using secret randomness.
If this randomness is not properly discarded or the party is malicious, the system's security is compromised, allowing false statements to be proven. 
Hence, ZCash \cite{zerocash} and other applications have moved away from circuit-specific trusted-setup protocols~\cite{trusted_set_up}.
On the other hand, Orion does not require a trusted setup or rely on computationally expensive elliptic curve cryptography, enabling a faster prover—but at the cost of large (8 MB) proofs.
To put this in perspective, Orion's proofs are roughly 4$\times$ larger than the maximum block size observed in Ethereum cryptocurrency~\cite{ethereum_blocksize_2024}. 
Thus, a private transaction with an Orion proof would not fit into a single Ethereum block.

HyperPlonk~\cite{hyperplonk} is a recent system gaining attention for the ZKP criteria it optimizes for, which are desirable in many applications.
It has small proofs (typically 5 KB), low verification complexity, 
and supports \textit{universal trusted setup}~\cite{universalsetup}, which runs once for all time and then can be reused by new applications.
A major contribution of HyperPlonk is its elimination of the Number Theoretic Transform (NTT), which is used by Groth16 and others, and instead its usage of SumCheck,
which lowers the asymptotic runtime from $O(n\log(n))$ (NTT) to $O(n)$ (SumCheck).
Here, $n$ represents the size of the underlying computation, and thus a lower runtime implies larger problems may be applied to ZKPs;
something we cannot do today due to long runtime.
Despite the improved runtime, the proving phase of HyperPlonk is still slow, running on the order of minutes to hours for some applications.

Accelerating HyperPlonk presents many challenges.
First, the prover works over large, computationally demanding data structures, with polynomials on the order of 2$^{17}$-2$^{24}$ and bitwidths ranging from 255-381.
Second, the protocol constitutes four major phases of computation, each comprising a different set of kernel functions.
Kernels range from small and large compute-bound problems (e.g., SHA3 and MSM, respectively) to memory-bound (e.g., SumCheck).
Compared to more traditionally accelerated kernels (e.g., NTT, matrix multiply, and convolution), HyperPlonk's kernels are also less well understood.
Finally, there is heterogeneity in data access patterns across kernels and phases.
Some data structures are reused, either between or within kernels, and some are not.
This raises questions on how to connect and schedule kernels as well as how to stage data to overlap communication with computation.
The complexity of HyperPlonk requires careful design, new hardware optimizations, and thorough design space exploration to identify architectures with area-performance tradeoffs that both meet the computational needs of applications and justify the costs of hardware acceleration.

In this paper we present zkSpeed, a modular HyperPlonk accelerator that overcomes the challenges outlined above.
zkSpeed comprises eight unique kernel accelerators, called units.
Each unit supports required operators (e.g., large-word modular arithmetic) natively in hardware over a programmable problem size.
The units collectively implement the complete zkSpeed chip architecture, which includes shared and unit-local scratchpads to capture reuse, capable of processing entire HyperPlonk proofs.
Units and scratchpads communicate via a multi-channel shared bus, and we propose a schedule of each (sequential) HyperPlonk protocol phase onto zkSpeed accordingly.
The schedule balances off-chip bandwidth constraints with on-chip reuse from scratchpads, unit computational throughput, and bus resource contention with a goal of overlapping off-chip communication with computation whenever possible.
All unit designs and bandwidth constraints are configurable and can be tailored to use the least amount of area needed to achieve this goal.
To optimize all units under a given off-chip bandwidth, we developed a zkSpeed simulator capable of sweeping over a wide range of design parameter settings.
We consider thousands of unique zkSpeed designs.
Analyzing the resulting Pareto frontier, we conclude that a 366mm$^2$ chip with 2 TB/s of off-chip bandwidth can speedup proof generation by 801$\times$ (geomean) over a CPU baseline.

This paper makes the following contributions:
\begin{itemize}
    \item A high-throughput, fully-pipelined accelerator to handle three flavors of SumCheck for HyperPlonk-based proofs.
    \item A novel implementation for modular inversion to compute fraction polynomials not seen in other ZKP protocols.
    \item Additional optimizations--including resource sharing of modular multipliers across units and on-chip memory compression schemes--that save upwards of 50\% area per unit and up to 85\% in bandwidth utilization.
    \item A comprehensive design space exploration of all hardware units to investigate design tradeoffs as we scale to high-performance designs and advanced memory technologies.
    \item A full-chip design that achieves 801$\times$ gmean speedup over CPU at iso-compute area.
\end{itemize}

\section{Background}
In this section, we describe state-of-the-art ZKPs and introduce the mathematical principles underlying the key kernels of HyperPlonk.

\subsection{zkSNARKs}
Today, the state-of-the-art ZKP protocols are zero-knowledge Succinct Non-interactive Arguments of Knowledge (zkSNARKs). zkSNARKs have three properties: (i) zero-knowledge, i.e., the proof does not reveal any information about the secret witness $w$; (ii) succinct, i.e., the proof has a few hundreds of bytes; and (iii) non-interactive, i.e., $\mathcal{P}$ sends the  proof to  $\mathcal{V}$ in one exchange.

zkSNARKs like HyperPlonk use polynomials to encode the correct execution of the target program. Polynomials can encode complex computations and constraints in a compact form, reducing the computational burden on $\mathcal{V}$ to a few checks, instead of requiring inspection of individual operations in the program. In HyperPlonk, the two most time-consuming kernels are SumCheck and Multi-Scalar Multiplication (MSM). The SumCheck kernel lets $\mathcal{P}$ demonstrate knowledge of a polynomial by verifying that its properties hold over a large set of inputs without revealing, e.g., the polynomial's coefficients. MSMs ensure that $\mathcal{P}$ is bound to that polynomial, preventing manipulation of the polynomial in the proof.

\subsection{The SumCheck Kernel} \label{sec: SumCheck}
SumCheck \cite{thaler_proofs_args_zk} is an interactive protocol between a prover $\mathcal{P}$ and a verifier $\mathcal{V}$. 
$\mathcal{P}$ demonstrates to $\mathcal{V}$ that it has correctly computed the sum of a polynomial over the \textit{boolean hypercube}, i.e., over all Boolean (0/1) assignments of its variables. 

Given a multivariate polynomial $P(x_1, x_2, \dots, x_\mu)$, where  $x_i \in \mathbb{F}_q$\footnote{each variable can be an integer modulo a prime number $q$.}, 
$\mathcal{P}$ wants to prove to $\mathcal{V}$ that it correctly
computed the sum 

\begin{equation*}
    H = \sum_{x_1 \in \{0,1\}} \sum_{x_2 \in \{0,1\}} \cdots \sum_{x_\mu \in \{0,1\}} P(x_1, x_2, \dots, x_\mu).
\end{equation*}

$\mathcal{P}$ and $\mathcal{V}$ engage in a multi-round protocol.
In Round 1, $\mathcal{P}$ computes $g_1(x_1)$, a \textit{univariate} polynomial of $x_1$ by summing over all binary values of the remaining variables:
\begin{equation*}
    g_1(x_1) = \sum_{x_2 \in \{0,1\}} \sum_{x_3 \in \{0,1\}} \cdots \sum_{x_\mu \in \{0,1\}} P(x_1, x_2, \dots, x_\mu)
\end{equation*}
$\mathcal{P}$ then sends coefficients or evaluations of $g_1(x_1)$ to $\mathcal{V}$. $\mathcal{V}$ checks that $g_1(0) + g_1(1)=H$, and if so, generates a random challenge $r_1\in \mathbb{F}_q$ and sends it to $\mathcal{P}$. 

In Round 2, $\mathcal{V}$ asks $\mathcal{P}$ to prove that 
\begin{equation*}
    g_1(r_1) = \sum_{x_2 \in \{0,1\}} \sum_{x_3 \in \{0,1\}} \cdots \sum_{x_\mu \in \{0,1\}} P(r_1, x_2, \dots, x_\mu), 
\end{equation*}
which is simply an instance of SumCheck, except over a ($\mu - 1$)-variate polynomial. 
Thus, in Round 2, $\mathcal{P}$ computes and returns
\begin{equation*}
    g_2(x_2) = \sum_{x_3 \in \{0,1\}} \sum_{x_4 \in \{0,1\}} \cdots \sum_{x_\mu \in \{0,1\}} P(r_1, x_2, \dots, x_\mu)
\end{equation*}
to $\mathcal{V}$ and the protocol repeats recursively for a total of $\mu$ rounds. If all checks pass, $\mathcal{V}$ accepts $\mathcal{P}$'s claim about $H$.

\subsection{Multilinear Polynomials}\label{sec:mle_poly}
ZKPs like HyperPlonk use \emph{multilinear polynomials}, that are \textit{linear} in each of their variables. For example, $$f(x_1, x_2, x_3) = x_1x_2 + 2x_1x_3 + 3x_1x_2x_3$$ is multilinear because the maximum degree of individual variables is one.
ZKP protocols use a general representation of multilinear polynomials, shown below for a 2-variable multilinear polynomial:
\begin{equation*}
    f(x_1, x_2) = (1-x_2)(1-x_1)a_0 + x_2(1-x_1)a_1 + (1 - x_2)x_1a_2 + x_2x_1a_3.
\end{equation*}
This representation defines polynomials using evaluations at specific input points (e.g., $f(1, 0) = a_2$) rather than using coefficients of symbolic terms (e.g., $x_1x_2$). Just as a degree-$d$ polynomial is uniquely determined by $d+1$ coefficients, it can also be uniquely defined by its values at $d+1$ distinct points.
This representation then maps cleanly to hardware in the form of lookup tables. For example, for a 2-variable multilinear polynomial, the binary values of $(x_1, x_2)$ index into the table to get the corresponding evaluation.
In general, a multilinear polynomial with $\mu$ variables $x_1 \ldots x_\mu$ can be stored in a lookup table of $2^\mu$ entries.
As we will discuss in Section \ref{sec: hyperplonk_protocol}, HyperPlonk uses multilinear polynomials as building blocks for higher-degree polynomials on which we then perform SumCheck and other computations. 
In the rest of the paper, we will use the term \textit{MLE table} to refer to these lookup data structures (MLE stands for ``multilinear extensions'' \cite{thaler_proofs_args_zk}; the word ``extension'' indicates that these polynomials can  be evaluated at non-boolean, or extended, values).

\subsection{The MSM Kernel}
MSMs are dot products between a vector of scalars $\vec s$ and a vector of 2D or 3D points $\vec P$ on an elliptical curve, e.g., $\sum_{i=0}^{n-1} s_{i}P_{i}$.
MSMs are used in ZKPs to perform \textit{commitments}. A commitment is a cryptographic primitive that binds a prover to a value without revealing it.
In zkSNARK protocols like HyperPlonk \cite{hyperplonk},  scalars are the polynomial evaluations stored in MLE tables.
Computing a dot-product reduces polynomials to a single value, i.e., the commitment.

MSMs are a bottleneck in ZKP provers and recent work has focused on accelerating them on ASIC and GPU \cite{szkp, pipezk, gzkp, priorMSM, distMSM, cuZK, reZK, gypso, myotosis, graz}.
The bottleneck is due to the extremely expensive elliptic curve point multiplications that they use. To reduce this cost, MSMs use  Pippenger's algorithm~\cite{pippenger}, which performs point multiplications via several point additions (PADDs). PADDs are still expensive, typically tens of regular modular multiplications. In the context of ASIC accelerators, the state-of-the-art (SZKP \cite{szkp}) presents a framework for building scalable MSM architectures.

\section{The HyperPlonk Protocol}\label{sec: hyperplonk_protocol}
In this section, we describe how HyperPlonk encodes the computation being proven, how this encoding shapes the structure of the underlying polynomials, and the key steps in proof generation.

\subsection{Plonk-based encodings} 
In ZKPs, the program being proven must be converted into a specific form before generating the proof. While most prior works use \textit{Rank-1 Constraint System (R1CS)} \cite{libsnark, groth}—a series of sparse matrix-vector encodings—HyperPlonk adopts a Plonk-based polynomial structure \cite{plonk}. R1CS and Plonk encodings map all program computations into addition, multiplication, and boolean operations, with nonlinear functions (e.g., branches) resolved via bit-wise decompositions.

Plonk-based encodings map each operation in a program's execution to a \textit{gate} -- similar in principle to gates seen in other privacy-preserving protocols \cite{haac, mo2025able, chillotti2020tfhe, 10.1145/3617232.3624852} -- that supports addition, multiplication, conditionality, and equality checks, as shown in \autoref{eq:gate}:
\begin{equation}
    f = q_Lw_1 + q_Rw_2 + q_Mw_1w_2 - q_Ow_3 + q_c
    \label{eq:gate}
\end{equation}

This can be viewed as a ZKP-specific instruction set where
each gate acts as an instruction performing basic arithmetic or logical operations.
In \autoref{eq:gate}, terms $q_{L}, q_{R}, q_{M}, q_{O}$ are binary \textit{control} signals for left, right, multiply, and output 
ports. 
The term $q_{c}$ represents a constant input. Terms $w_{1}, w_{2}$ represent the gate's inputs, and $w_{3}$ is its output. These last three terms are collectively called \textit{witnesses}. 
An addition operation, for instance, is implemented by setting 
$q_{L}=1$, $q_{R}=1$, $q_{O}=1$ and other control and constant inputs to $0$. 
This yields: $f = w_1 + w_2 - w_3$. That is, $f=0$ if and only if the 
addition is correctly performed.

Each operation in a program is mapped to a Plonk gate. A program
with $2^{\mu}$ operations has $2^{\mu}$ Plonk gates connected to 
form a \textit{circuit}.  A $\mu$-variate polynomial $f(X_{1}, X_{2} \ldots, X_{\mu})$ represents the entire circuit, with 
binary assignments to variables $\{X_{1}, X_{2} \ldots, X_{\mu}\}$ 
representing individual gates. 
For example, if $\mu = 4$, the circuit has $16$ gates; $f(X_{1}=0, X_{2}=0, X_{3} = 0, X_{4}=0)$ represents the 
first gate.

$f(X)$ is constructed using \textit{multilinear} polynomials representing every term from Eq. \ref{eq:gate}.
For example, $q_{L}(X)$ is the left control input polynomial, $w_{1}(X)$ is the first data input polynomial, and so on.
Each of these polynomials also takes $X = [X_1, X_2, \ldots, X_{4}]$ as an index.
For example, in a program that maps to 16 gates, $q_M([0, 1, 0, 0])$ will tell us whether or not the multiplication control signal is enabled in gate 2. 
$w_3([1, 1, 1, 1])$ tells us what the output data signal is at gate 15.
All individual polynomials are encoded in this fashion, and can be stored as MLE tables.
These MLE tables are populated from the program trace in software before running the HyperPlonk prover.
The MLEs are then used by the steps of the HyperPlonk protocol, which we discuss in the following sections.

\subsection{SumCheck on Plonk-based Encodings}   
\begin{figure}
  \centering
  \includegraphics[width=\columnwidth]{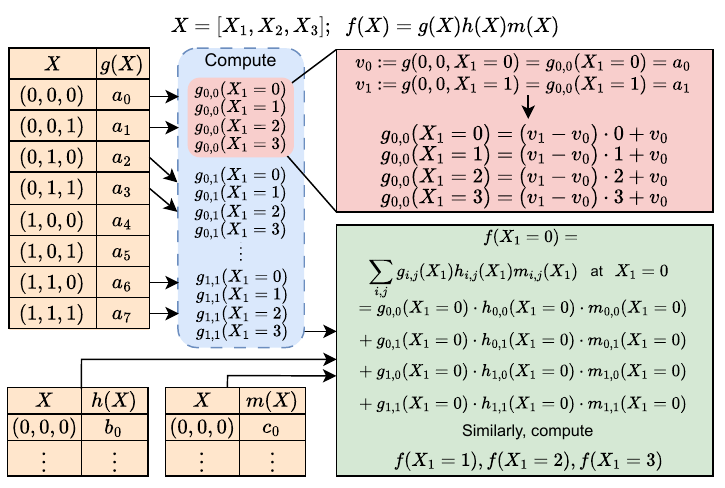}
  \vspace{-1em}
  \caption{SumCheck Example. The subscripts $(0,0)$, $(0,1)$,  etc. refer to the specific boolean instances of $X_2, X_3$.}
  \Description{SumCheck example illustrating the computation for an example polynomial.}
  \label{fig:SumCheck_example}
  \vspace{-1em}
\end{figure}

The key difference between the SumCheck used in HyperPlonk and the example in Section~\ref{sec: SumCheck} lies in the structure of the underlying polynomials. While the earlier example operates on a single polynomial of arbitrary degree, HyperPlonk applies SumCheck to a version of the Plonk polynomial in~\autoref{eq:gate}, as well as other polynomials discussed in~\autoref{sec:sc_hw_section}.
These polynomials consist of the \textit{products} of multilinear polynomials, with terms of varying degrees.
\autoref{fig:SumCheck_example} shows an example of how a SumCheck of the product of three multilinear polynomials is computed. 

In the example from Section \ref{sec: SumCheck}, if $g(X)$ was multilinear, we would only need the evaluations at $X_1 = 0$ and $1$ because the result of summing over all variables yielded a univariate degree-1 polynomial.
In this example, however, after summing over all variables, we are left with a univariate \textit{degree-3} polynomial.
As mentioned in \autoref{sec:mle_poly} (and from basic polynomial principles), a degree-3 polynomial must be evaluated at \textit{4} unique points to fully characterize the polynomial.
In the figure example, in the blue region, we are iterating over $[X_2,X_3] = [0,0], [0, 1], [1, 0], [1, 1]$ to compute the evaluations. 
For each iteration (red region), we must therefore evaluate each polynomial at $X_1 = 0, 1, 2, 3$. (Here, $X_1 = 2, 3$ are the \textit{extensions} that we compute from the existing evaluations at $0, 1$.)
Then, in the green region, we compute the product for $X_1 = 0$ across all polynomials, and then sum across iterations. This is repeated for $X_1 = 1, 2, 3$.
After the summations, each polynomial's MLE table is updated with a random challenge $r$ from the verifier. We refer to this step as \textit{MLE Update}. For every boolean hypercube instance (e.g., $(X_2, X_3)$ in \autoref{fig:SumCheck_example}), we extrapolate the linear polynomial in $X_1$ to the value of $r$. 
For example, a table $t'$ (for round 2) can be constructed from a table $t$ (from round 1) using the formula
\begin{equation}
    t'[i] \leftarrow (t[2i + 1] - t[2i])r + t[2i]
\end{equation}

The sum of all of these core computational steps is more expensive than baseline SumCheck.
An additional layer of complexity is that HyperPlonk's SumCheck polynomials exhibit degree variation across terms.
This leads to imbalance in how many total evaluations are needed per term (i.e., $q_Lw_1$ needs 3 evaluations, while $q_Mw_1w_2$ needs 4). This is handled by the HyperPlonk SumCheck protocol with a fixed interpolation step before the MLE tables are updated.
While these computations are expensive, many of these operations can be performed in parallel. All polynomials can compute their extensions independently, and within a polynomial, extensions at different values can be parallelized. Thus, HyperPlonk's SumChecks have high compute and degrees of parallelism.

\begin{figure*}[t]
\centerline{\includegraphics[width=1.03\textwidth]{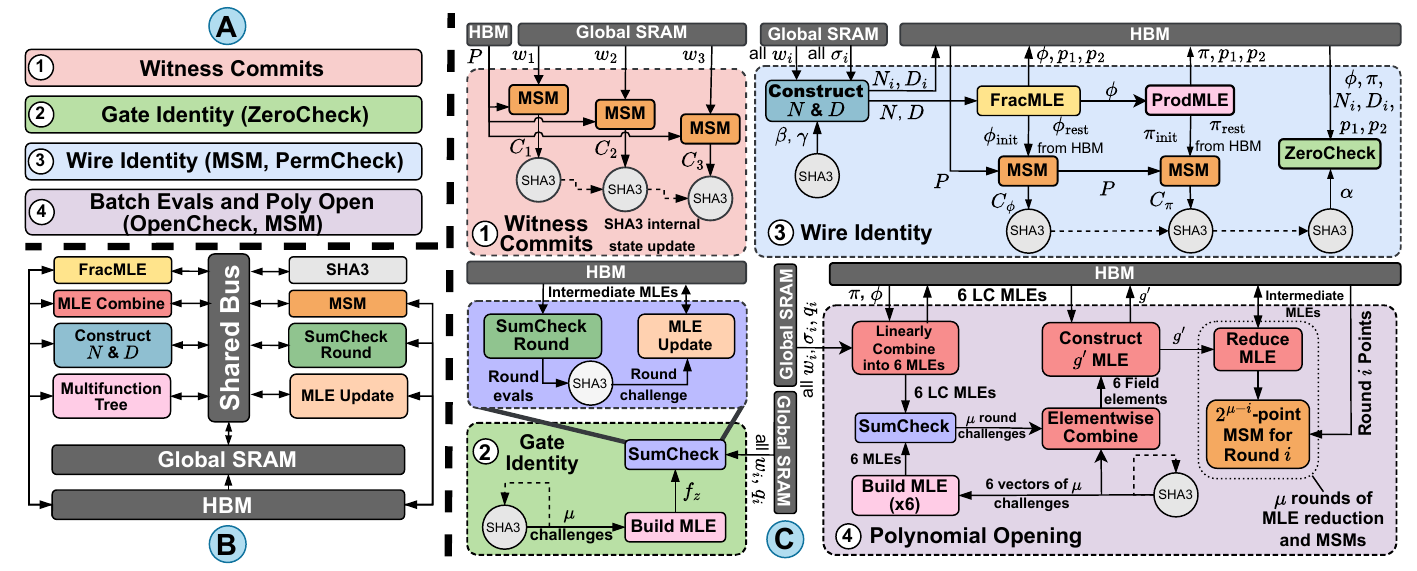}}
\vspace{-1em}
\caption{A) HyperPlonk Protocol Steps, B) zkSpeed Architecture, and C) Step Dataflow. Each step is executed in the numerical order listed. Colors of each module in the zkSpeed architecture directly correspond with the operations in the dataflow. In Wire Identity, PermCheck refers to the creation of MLEs and then the ZeroCheck that takes them as input. ProdMLE refers to the construction of Product MLE. Batch Evals only use the Multifunction Tree unit, they are omitted for space.}
\Description{HyperPlonk's protocol steps, zKSpeed's architecture, and dataflow for each protocol step.}
\label{fig:dataflow_and_arch}
\end{figure*}

\begin{figure*}[t]
\centerline{\includegraphics[width=2\columnwidth]{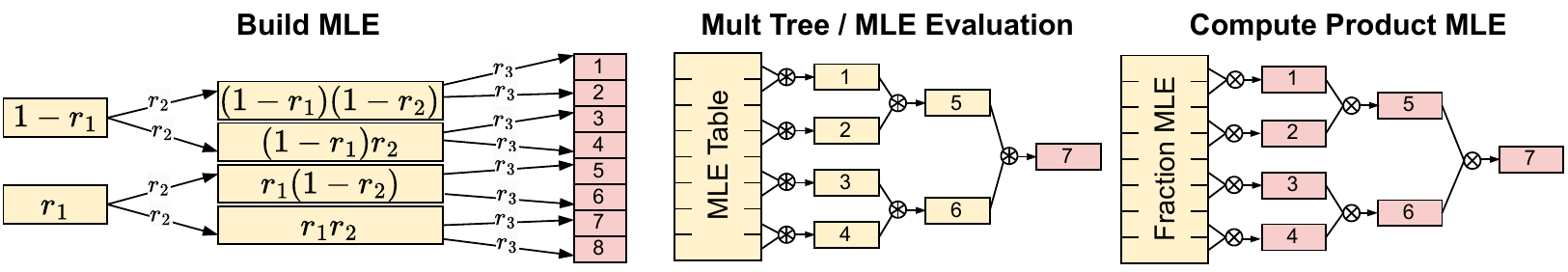}}
\vspace{-1em}
\caption{Different computation patterns that fit in the Multifunction Tree unit.
Computation outputs are marked in red. }
\Description{Scenarios that can be done with Multifunction Tree.}
\label{fig:tree_scenario}
\end{figure*}

\subsection{HyperPlonk Protocol Steps} 
\label{sec:HyperPlonk_Protocol_Steps}
In this section, we outline the steps of the HyperPlonk protocol, as seen in \autoref{fig:dataflow_and_arch}, what each step achieves, and what hardware units are needed to perform each step.

\subsubsection{Witness Commits} 
The first step in HyperPlonk's prover involves witness commitments.
In this step, we compute MSMs between elliptic curve points and the witness polynomials $w_1, w_2, w_3$ to reduce each polynomial to individual commitments.
The witness polynomials in HyperPlonk are typically ``Sparse'', meaning that 90\% of the values are either 0 or 1, and 10\% are up to the full scalar bitwidth. SZKP \cite{szkp} handles this by using \textit{Sparse MSMs}. We implement Sparse MSMs using the MSM unit for this step.

\subsubsection{Gate Identity}\label{sec:ZeroCheck}  
The Gate Identity step confirms that each gate in the circuit performs its computation correctly.
The intuition is that if so, then $f(X)$ in \autoref{eq:gate} should evaluate to 0 for each gate, or equivalently, for each binary assignment of $X$.
Further, the \textit{sum} of $f(X)$ over all $X$'s should also equal 0, which can be confirmed by running SumCheck on $f(X)$. 
However, since the prior statement is only a necessary but not sufficient condition for correct computation,
HyperPlonk performs SumCheck over a polynomial $f(X)r(X)$, where $r(X)$ is a multilinear polynomial with evaluations computed from $\mu$ random challenges 
(the $eq$ polynomial of Lemma 2.1 in \cite{hyperplonk}).
We will henceforth refer to the construction of such a polynomial as \textit{Build MLE}.
Then, the sum over the boolean hypercube of $f(X)r(X)$ should also be 0; this is known as a ZeroCheck.
The Gate Identity step relies on three units: the Multifunction Tree unit (Section \ref{sec:multifunction_tree}) to build $r(X)$, the SumCheck unit, and the MLE Update unit to update MLE tables between SumCheck rounds.

\subsubsection{Wiring Identity}\label{sec:permcheck}  
This step verifies that the outputs of each gate are routed correctly to inputs of downstream gates. 
This is achieved via a PermutationCheck (PermCheck), which is performed by  
constructing a numerator polynomial $N$ that encodes the natural ordering of the outputs and a denominator polynomial $D$ that encodes the permuted ordering (derived from permutation polynomials $\sigma_{1\ldots3}$), and then checking that $\phi = N/D$, which we refer to as Fraction MLE, equals 1. 
As we will discuss, computing $D^{-1}$ is time consuming, since inverting (or, in effect, performing divisions on) elements of a finite field is
difficult.
The Product MLE, $\pi$, is generated by computing a cumulative product over $\phi$ to form the first half, then recursively applying cumulative products on $\pi$ itself to complete the second half.
Then, the $\phi$ and $\pi$ polynomials are both \textit{committed} to via MSM computations. 
A final ZeroCheck is also performed to ensure that the structure imposed upon $\phi$ and $\pi$ was not violated during their creation, i.e., the sum over the boolean hypercube of the resulting constraint polynomial should be zero everywhere. All polynomials are stored in the form of MLE tables.
The Wiring Identity step uses the Construct N\&D unit, the FracMLE unit, the Multifunction Tree unit (to construct the Product MLE $\pi$ and later for the Build MLE step within ZeroCheck), the MSM unit, the SumCheck unit, and the MLE Update unit.

\subsubsection{Batch Evaluations} \label{sec: batch_eval}  
The SumCheck protocols require evaluating several MLEs at specific points.
This is achieved with batch evaluations, where the input MLEs, as well as $\phi$ and $\pi$, are queried at 6 points, some of which are derived from round challenges in the SumCheck portion of the Gate and Wiring Identity steps, and some of which are fixed at compile time.
In total, 22 total evaluations are performed among 13 polynomials using 6 distinct points.
This step is almost the exact reverse of the step in \ref{sec:ZeroCheck} used to construct $r(X)$, since we are compressing an entire MLE into 1 value, while $r(X)$ is an entire MLE built from $\mu$ values. We will henceforth refer to the operation of evaluating an MLE at a point (consisting of $\mu$ field elements) as \textit{MLE Evaluate}.
These polynomial queries are then used by the verifier to check that the prover obeyed the protocol.
The Batch Evaluation step uses the Multifunction Tree unit.

\subsubsection{Polynomial Opening}
\label{sec:polyopen}
This step succinctly verifies the correctness of the prover's batch evaluations. For brevity, we will omit details of the polynomial opening step. To summarize, it involves first computing 6 MLEs as a linear combination of the  MLEs from Eq. \ref{eq:gate} and $\phi$ and $\pi$. Then, 6 more MLEs are constructed from the query points in Section \ref{sec: batch_eval}. These 12 MLEs are linearly combined into a single MLE, upon which a final SumCheck is computed. To avoid confusion, we will refer to this final SumCheck as \textit{OpenCheck} and use \textit{SumCheck} to refer to the underlying computation.

After OpenCheck, the first 6 MLEs used as OpenCheck's inputs are linearly combined with the OpenCheck's round challenges to construct a final MLE, which is denoted as $g'$. This MLE is first reduced to half its size, and it is used as the scalar set for a $2^{\mu-1}$-point MSM. We then  halve the scalar set and perform MSMs. For example, if $\mu = 10$, we compute a $2^9$-point MSM, then a $2^8$-point MSM, all the way to a $2^0$-point MSM.
The Polynomial Opening step uses the MLE Combine unit to compute the linear MLE combinations, the Multifunction Tree unit to build the 6 MLEs, the SumCheck unit to perform OpenCheck, and the MSM Unit to compute MSMs.

\subsubsection{SHA3}
zkSNARKs are non-interactive and use SHA3 to generate challenges as well as maintain transcripts.
Transcripts are logs of proof-related computations checked by the verifier.
In HyperPlonk, SHA3 is invoked between steps to ``log'' computed values, which is achieved by updating the SHA3 state.
Intuitively, this ensures that all future challenges are bound to the existing history of computations logged in the transcript.
As a result, SHA3 effectively acts as an order-enforcing mechanism.
This means the protocol steps must be executed in series, as shown in \autoref{fig:dataflow_and_arch}.

\begin{table}[]
\centering
\caption{Modular multiplications, memory requirements, and arithmetic intensity of select functions for $2^{20}$ gates. Links to the source code are provided\tablefootnote{SumCheck code (used by ZeroCheck, PermCheck, and OpenCheck) is at \\
\urlicon{https://github.com/EspressoSystems/hyperplonk/blob/main/subroutines/src/poly_iop/sum_check/mod.rs\#L154}{ (lines 154-182)} and \urlicon{https://github.com/EspressoSystems/hyperplonk/blob/main/subroutines/src/poly_iop/sum_check/prover.rs\#L123}{ (lines 123 - 181)}}.
}
\vspace{-1em}
\label{tab:Arithmetic Intensity}
\resizebox{\columnwidth}{!}{
\setlength{\tabcolsep}{1mm}{
\begin{tabular}{@{}lccccc@{}}
\toprule
 \textbf{Kernel} & \begin{tabular}[c]{@{}c@{}}Source\\ Code\end{tabular} & \begin{tabular}[c]{@{}c@{}}Modmuls\\ (millions)\end{tabular} & \begin{tabular}[c]{@{}c@{}}Input\\ Size (MB)\end{tabular} & \begin{tabular}[c]{@{}c@{}}Output\\ Size (MB)\end{tabular} & \begin{tabular}[c]{@{}c@{}}Arithmetic Intensity\\ (modmul/byte)\end{tabular} \\ \midrule
\textbf{Poly Open MSMs} & \urlicon{https://github.com/EspressoSystems/hyperplonk/blob/main/subroutines/src/pcs/multilinear_kzg/mod.rs\#L254} & 1160 & 127 & 0.00     & 8.70 \\ \hline
\textbf{Wire Identity MSMs} & \urlicon{https://github.com/EspressoSystems/hyperplonk/blob/main/subroutines/src/poly_iop/prod_check/mod.rs\#L198} & 2290 & 254 & 0.00  & 8.59 \\ \hline
\textbf{Witness MSMs}       & \urlicon{https://github.com/EspressoSystems/hyperplonk/blob/main/hyperplonk/src/snark.rs\#L186} & 1370 & 167 & 0.00  & 7.83 \\ \hline
\textbf{Batch Evaluations}  & \urlicon{https://github.com/EspressoSystems/hyperplonk/blob/main/hyperplonk/src/snark.rs\#L279} & 23.1 & 77.5 & 0.00 & 0.28 \\ \hline
\textbf{ZeroCheck Rounds}   & \urlicon{https://github.com/EspressoSystems/hyperplonk/blob/main/subroutines/src/poly_iop/zero_check/mod.rs\#L69} & 77.6 & 332 & 0.00  & 0.22 \\ \hline
\textbf{Fraction MLE}            & \urlicon{https://github.com/EspressoSystems/hyperplonk/blob/main/subroutines/src/poly_iop/prod_check/util.rs\#L22} & 5.19 & 0.00 & 31.9 & 0.16 \\ \hline
\textbf{PermCheck Rounds}   & \urlicon{https://github.com/EspressoSystems/hyperplonk/blob/main/subroutines/src/poly_iop/prod_check/util.rs\#L122} & 94.4 & 701 & 0.00  & 0.13 \\ \hline
\textbf{Linear Combine}     & \urlicon{https://github.com/EspressoSystems/hyperplonk/blob/main/subroutines/src/pcs/multilinear_kzg/batching.rs\#L95} & 18.9 & 77.5 & 191  & 0.07 \\ \hline
\textbf{OpenCheck Rounds}   & \urlicon{https://github.com/EspressoSystems/hyperplonk/blob/main/subroutines/src/pcs/multilinear_kzg/batching.rs\#L128} & 31.5 & 765 & 0.00  & 0.04 \\ \hline
\textbf{Construct N \& D}      & \urlicon{https://github.com/EspressoSystems/hyperplonk/blob/main/subroutines/src/poly_iop/perm_check/util.rs\#L30} & 10.5 & 18.2 & 255  & 0.04 \\ \hline
\textbf{Product MLE}      & \urlicon{https://github.com/EspressoSystems/hyperplonk/blob/dc194f83ef5cae523b869f7256f314bdbeb2a42c/subroutines/src/poly_iop/prod_check/util.rs\#L65} & 1.05 & 0.00 & 31.9  & 0.03 \\ \hline
\textbf{All MLE Updates}      & \urlicon{https://github.com/EspressoSystems/hyperplonk/blob/main/subroutines/src/poly_iop/sum_check/prover.rs\#L91} & 33.6 & 1800 & 900 & 0.01 \\ \bottomrule
\end{tabular}
}}
\end{table}

\subsubsection{Compute Demands of HyperPlonk}\label{sec:Compute_Demands}
Table~\ref{tab:Arithmetic Intensity} summarizes profiling results to characterize key functions and understand the sources of performance overhead and hardware needs when accelerating HyperPlonk.
There are too many functions to list, so we present the twelve with the highest arithmetic intensity, which is defined as modular multiplications (modmuls) per byte, as done in prior work~\cite{ark, does_fhe_need_accelerators} (note SHA3 has no modmuls).
Additionally, the reference CPU implementation is provided as a link.
First, we observe that all functions require an immense amount of computation: ranging from millions to billions of 255/381-bit modmuls (comprising three integer multiplications) over all invocations of each function.
For example, the data for Wire Identity MSMs modmuls reflects two function calls; for ZeroCheck Rounds there is one function call (see links in table).
This motivates the need for both specialized modmul units and a high degree of parallelism to mitigate overhead.
Second, compute intensity drops off sharply after the third function (since data reuse is limited), and the data input/output sizes for all functions are large, typically hundreds of megabytes up to terabytes as problem sizes scale to more gates. This motivates the need for large on-chip scratchpads to mitigate off-chip data movement when possible and high off-chip (i.e., HBM) bandwidth.

\section{HyperPlonk Accelerator Units}
\label{sec:HyperPlonk_Accelerator_Units}
HyperPlonk's protocol is based on the BLS12-381 elliptical curve. Here, all MLE datatypes are 255 bits wide and all elliptical curve points in the MSMs are 381 bits wide. All MLE and MSM operations involve modular arithmetic primitives. These are built into each accelerator unit that requires them.
zkSpeed comprises eight accelerator units and we describe each below.

\begin{figure}
  \centering
  \includegraphics[width=0.5\textwidth]{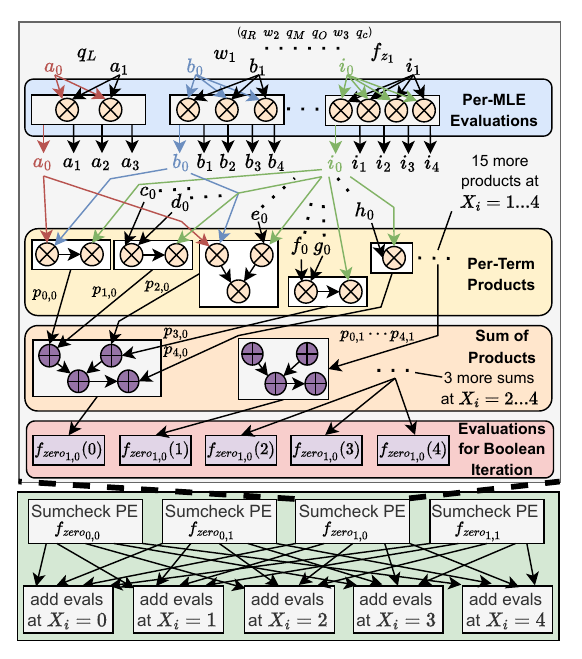}
  \caption{SumCheck Round unit example. Subscripts of $f$ indicate which $X_2, X_3$ instance is being handled by the PE. For simplicity, each MLE is renamed from $a-i$. Their subscripts indicate the evaluation, e.g., $X_1 = 0\ldots4$. The products $p_{j,k}$ are labeled to indicate the $j^{th}$ term evaluated at $X_1 = k$.}
  \label{fig:sumcheck_uarch}
\end{figure}

\subsection{SumCheck and MLE Update}\label{sec:sc_hw_section}
Three HyperPlonk steps use SumCheck: ZeroCheck (Section \ref{sec:ZeroCheck}), PermCheck (Section \ref{sec:permcheck}), and OpenCheck (Section \ref{sec:polyopen}). 
Each step has a unique polynomial shown, respectively, below: 
\begin{equation}\label{eq:zc}
    f_{zero} = q_Lw_1f_{z_1} + q_Rw_2f_{z_1} + q_Mw_1w_2f_{z_1} - q_Ow_3f_{z_1} + q_cf_{z_1}
\end{equation}
\begin{equation}\label{eq:pc}
    f_{perm} = \pi f_{z_2} - p_1p_2f_{z_2} + \alpha(\phi D_1D_2D_3 )f_{z_2} - \alpha (N_1N_2N_3)f_{z_2}
\end{equation}
\begin{equation}\label{eq:oc}
    f_{open} = y_1k_1 + y_2k_2 + y_3k_3 + y_4k_4 + y_5k_5 + y_6k_6
\end{equation}

In these equations, $\alpha$ is a challenge from the SHA3 unit, and all other symbols represent multilinear polynomials.
$f_{z_1}, f_{z_2}$ and $k_i$ are all \textit{Built MLEs} constructed from SHA3 challenges (see the discussion of $r(X)$ in \autoref{sec:ZeroCheck}).
These SumCheck polynomials share a common sum-of-products representation, but are each unique and require slightly different datapaths.
We develop a unified SumCheck unit that can handle each of these polynomials and
highlight the key contributions of our architecture next.

\subsubsection{SumCheck Round PE Microarchitecture}
\label{sec:round_pe}
In ZeroCheck and PermCheck, there are polynomials that appear multiple times across terms.
In HyperPlonk's CPU baseline, the boolean hypercube summations are performed iteratively term-by-term, incurring redundant computation for these repeating polynomials. 
We address this by computing all evaluations for each polynomial in parallel. For example, in Equation \ref{eq:zc}, the polynomial $f_{z_1}$ has to be evaluated at $X_i = 0, 1, 2, 3$ and $4$ because the largest-degree term it appears in has a degree of four. This computation only needs to be performed once before being used to compute the product in each term of $f_{zero}$. 
\autoref{fig:sumcheck_uarch} shows an example of how our SumCheck PEs handle ZeroCheck computation for $f_{zero}$. Each unique polynomial takes its respective $X_i = 0 \text{ and } 1$ values (e.g., $c_0 = q_R(X_i = 0)$ in \autoref{fig:sumcheck_uarch}), and uses these values to extend to the needed $X_i$ evaluations. Then, each product in \autoref{eq:zc} must be computed for each $X_i$ evaluation. The products at the evaluation points are then accumulated into registers. 
Due to the inherent degree imbalance, some terms have fewer evaluations; the additional evaluations are computed via Barycentric Interpolation \cite{barycentric}.
This is omitted in the figure since it only adds a fixed cost at the end of each round (23 modmuls for ZeroCheck and 46 modmuls for PermCheck).
Each polynomial observes a different datapath, so we use a specialized design to exploit high reuse, full-pipelining, and high levels of parallelism.

\subsubsection{Streaming approach}
At the start of the protocol, the MLE tables corresponding to polynomials in \autoref{eq:gate} are all provided to the prover and can be stored on-chip.
However, as these MLEs undergo rounds of SumCheck, the process of incorporating challenges into the MLE tables expands binary values to full 255-bit values.
Though the number of MLE table entries reduces by half each round, the data itself grows by over 100$\times$ between rounds 1 and 2, so the total storage cost for storing all MLEs is intractable.
However, each round, the intermediate values of MLE tables are only used by the main SumCheck computation and then by the MLE Update unit to be halved in size.
Since there is no data reuse in-between rounds, we adopt a streaming-based solution to alleviate the pressure on on-chip SRAM storage.
The key tradeoff here is that our SumCheck and MLE Update units become memory-bound, since each MLE in Equations \ref{eq:zc}-\ref{eq:oc} must be updated and written back to off-chip memory after SumCheck rounds, necessitating off-chip traffic.
Fortunately, recent advances in high-bandwidth memory (HBM) can supply very high bandwidths to offset this.
This is inline with many other cryptographic computing accelerators, which also rely on HBM~\cite{szkp, osiris, f1, clake, ark, sharp, bts}.
We analyze the bandwidth sensitivity of the SumCheck computations in our evaluation section.

\subsubsection{Scaling to Multiple PEs}
\label{sec:multi_pe}
Each SumCheck PE handles the product and sum for one iteration over the boolean hypercube. For example, in \autoref{fig:SumCheck_example}, this corresponds to one red region per polynomial (MLE table) and the first product in the green region for \textit{each} $X_1 = 0, 1, 2, 3$. These iterations run in parallel with multiple PEs storing their own accumulation registers. The bottom of \autoref{fig:sumcheck_uarch} shows the accumulation of each evaluation across boolean hypercube values (equivalently, across indices of the MLE tables). After all evaluations are done, an MLE Update PE handles the updates for one MLE table and can provision multiple modmuls. Multiple PEs can run in parallel, handling MLE tables independently.

\subsubsection{Unified SumCheck PE}
\label{sec:Unified_SumCheck_PE}
The three polynomials $f_{zero}$, $f_{perm}$, $f_{open}$ have different datapaths.
We use HLS tools to generate a unified PE that handles each SumCheck variation used in HyperPlonk. Each PE requires 94 modular multipliers, compared to 184 modular multipliers without resource sharing, saving 48.9\% on area. 

\subsubsection{Comparison to Prior SumCheck Implementations}
HyperPlonk's CPU library is designed to support \textit{any} composition of multilinear polynomials for different protocols, not just the three we have shown. 
Consequently, repeating polynomial computations as observed in the CPU implementation greatly improves programmability, as opposed to having specialized functions to handle the specific computation patterns we optimize for.

NoCap~\cite{nocap} is a recent accelerator that also accelerates SumCheck, but there are critical differences at the protocol-level that motivate our architecture. 
NoCap implements Spartan~\cite{spartan}, which uses R1CS encodings resulting in two SumCheck instances that look as follows: $f_{1} = g_1g_2g_4 - g_3g_4$ and $f_{2} = g_5g_6$.
NoCap uses a vector architecture with 2048 PEs to process boolean hypercube instances with a Beneš network to sum across PEs.
This makes sense for NoCap because Spartan's polynomials are degree $2$ and $3$ with up to two terms. 
In contrast, HyperPlonk's polynomials have a more heterogeneous structure; there are more terms of varying degree. 
This complexity arises from the usage of the control and constant polynomials ($q_{L}, q_{R}, q_{M}, q_{O}, q_{C}$) to represent gates. 
These are required to keep HyperPlonk verifier costs low (in Section~\ref{sec:Evaluation}, we see NoCap's verifier is slower). 
Consequently, mapping HyperPlonk's polynomials to a vector architecture would put more pressure on vector register files because of the high amount of intermediate values needed to be read. 
As seen in \autoref{fig:sumcheck_uarch}, our SumChecks also require very complex communication, which can increase bandwidth pressure and may not be efficient to implement with a Beneš network.
Further, our specialized PEs immediately reuse values without relying on register files to store the numerous amount of intermediates. We have fewer, heftier PEs, so data movement costs are less relative to those incurred by a vector processor.

\subsection{MSMs}
\label{sec:msm_para}

\begin{figure}[t!]
\centerline{\includegraphics[width=\columnwidth]{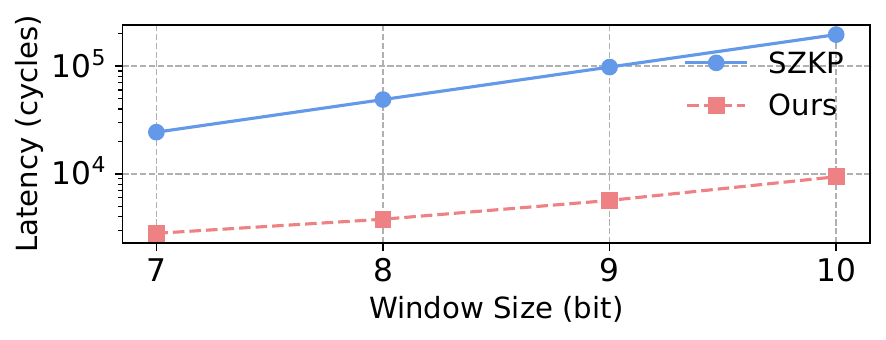}}
\caption{MSM bucket aggregation comparison.}
\vspace{-1em}
\Description{PADD bucket.}
\label{fig:PADD_bucket}
\end{figure}

MSMs are used in three HyperPlonk steps: Witness Commits, Wiring Identity, and Polynomial Opening.
Several prior works accelerated MSMs; we begin with the base MSM design from SZKP \cite{szkp}, the state-of-the-art for accelerating Groth16, and propose two optimizations for better performance and area efficiency.

In HyperPlonk, the Sparse MSMs run in series and are on the critical path.
This is unlike prior protocols like Groth16 where their execution could be masked via parallel processing with the Dense MSMs.
Consequently, we opt to use the same MSM hardware unit for both Sparse and Dense MSMs in zkSpeed (as opposed to separate units in SZKP).
zkSpeed uses a similar scheme as SZKP for handling sparse computations.
First, we compute the sum of all points corresponding to $1$-valued scalars. This is done by fetching the points corresponding to $1$-valued scalars into the MSM unit's SRAM banks.
Then, using a tree-based approach, we feed two points at a time to the pipelined Point Adder (PADD) unit, with the result written back to the SRAM banks. This process repeats until we have reduced all points to the final sum for points corresponding to $1$-valued scalars. Note that in this step, we need not fetch scalars into the MSM's SRAM banks since they are all $1$.
Then, we use Pippenger's algorithm \cite{pippenger} on the remaining $\approx$ 10\% of (dense) scalars.
For the Dense MSMs in the Wiring Identity and Polynomial Opening steps, we use Pippenger's algorithm for the full MSM computation. We now discuss the two improvements to SZKP's MSM architecture.

\subsubsection{Reduced Memory Footprint} First, we note that elliptical curve points, while being three-dimensional, are initialized as $(X, Y$, 1) coordinates in HyperPlonk. 
Thus, we only fetch two coordinates per point, saving off-chip bandwidth.
We further save on-chip SRAM area.
SZKP provisions one scalar memory bank and three point memory banks, to hold $X, Y, Z$ coordinates.
While Sparse MSMs need not store the 1-valued scalars, the tree-based addition partials still need buffering (their $Z$ coordinates are no longer 1-valued) necessitating the $Z$ coordinate memory bank.
Therefore, zkSpeed allocates three SRAM memory banks.
In Dense MSM operation, the $Z$ memory bank is reused to store the scalars, and since partial sums are \textit{only} stored in bucket registers, we do not need to provision another dedicated memory bank, as in SZKP. This represents a savings of 18\% in on-chip SRAM area compared to having a dedicated scalar memory bank.

\subsubsection{Faster Bucket Aggregation} The second optimization addresses a runtime bottleneck in SZKP's bucket aggregation step.
After sorting points into buckets and computing each bucket's sum, SZKP employs a naive aggregation algorithm to calculate the sum $\sum_{i=1}^{2^W - 1} iB_i $, where $W$ is the window size \cite{szkp, pippenger}, and $B_i$ represents the accumulated sum of the $i$-th bucket.
This is inefficient when processing smaller MSMs, e.g., 32-point MSMs, which are prominent in Polynomial Opening.
The fixed bucket aggregation latency becomes a performance bottleneck because the point additions are serially performed and do not leverage the pipelining available in the PADD unit. Consequently, the PADD is underutilized in this step.
To address this, zkSpeed adapts bucket aggregation introduced in \cite{priorMSM}.
This scheme divides aggregation into smaller groups, computes the partial sums within each group in parallel, and finally combines the results.
As shown in Figure~\ref{fig:PADD_bucket}, it reduces the bucket aggregation latency by an average of 92\% across all window sizes compared to SZKP.
We select a group size of 16, which provides the best overall performance and ensures the aggregation step no longer dominates runtime for small MSMs.

\subsection{Multifunction Tree Unit}
\label{sec:multifunction_tree}
\subsubsection{Tree-like Computation in HyperPlonk}
Many HyperPlonk functions exhibit binary-tree compute patterns, including Build MLE, 
MLE Evaluate, and constructing Product MLE ($\pi$).
zkSpeed supports these in hardware with our Multifunction Tree Unit (MTU), specially designed to handle these compute patterns effectively.
Build MLE is a function used in ZeroCheck and OpenCheck steps.
It constructs a table with $2^\mu$ entries from random values $r_1$ to $r_\mu$, where $2^\mu$ represents problem size.
The computation is divided into $\mu-1$ layers (as a binary tree) to reduce the number of modular multiplications from $(\mu-1)2^{\mu}$ to $2^{\mu+1}-4$.
The multiplier tree is used for batch inversion during Fraction MLE generation (see \autoref{sec:gen_nd}), where it efficiently calculates the product $D[0] D[1] \cdots D[b-1]$ for an inversion batch size, $b$.
Similarly, the MLE Evaluate in the Batch Evaluation step operates like a multiplier tree but includes additional modular additions in each operation.
The Product MLE generation in the Wiring Identity outputs all layer results.
The functions' dataflows are presented in \autoref{fig:tree_scenario}.

\subsubsection{zkSpeed's Tree Approach}
The original (CPU) HyperPlonk implementation uses breadth-first (level-order) traversal (BFS).
This is inappropriate for hardware acceleration as it puts increased pressure on SRAM capacity and off-chip bandwidth.
For example, a problem size of $2^{23}$ requires up to $2^{22}$ intermediate elements in a
level, each 255 bits wide, which would require 128MB for the intermediates alone.
We propose a hybrid strategy that applies depth-first traversal (DFS) to the upper levels of the binary tree and switches to BFS at lower levels. 
This approach leverages efficient memory reuse in the upper levels through DFS, while enabling parallelism in the lower levels via BFS.
It reuses and consumes intermediate results as they are produced, reducing 
the amount of data that must be stored on-chip or 
spilled to DRAM.

\begin{figure}[t]
\centerline{\includegraphics[width=\columnwidth]{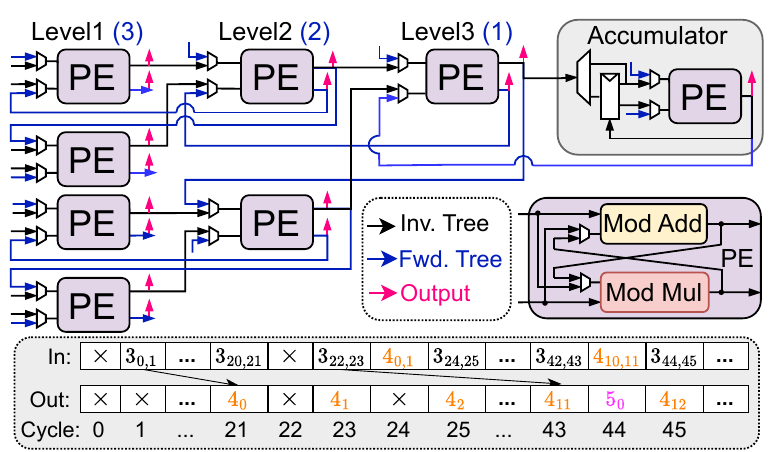}}
\caption{Hardware structure of the Multifunction Tree unit and accumulator schedule
(blue: forward, black: inverted, red: outputs to HBM/other modules. Level for binary tree).
Accumulator schedule shows the level and node index.
}
\vspace{-1.5em}
\Description{Hardware structure of the Multifunction Tree unit.}
\label{fig:tree_arch}
\end{figure}

\subsubsection{Multifunction Tree Architecture}
We further analyze this tree architecture in \cite{mulfunc_tree}.
Figure~\ref{fig:tree_arch} shows an example of how the MTU works.
Each PE includes a modular multiplier and modular adder\footnote{For Build MLE, $(1-r_1)r_2$ and $(1-r_1)(1-r_2)$ only require one modular multiplication since $(1-r_1)(1-r_2)$ is computed as $(1-r_1)-(1-r_1)r_2$}.
The hardware supports three processing modes of Figure~\ref{fig:tree_scenario}.
For the Inverse Tree (e.g., MLE Evaluate), the unit accepts $p$ inputs in parallel and reduces tree-level partials via a matching hardware dataflow; in Figure~\ref{fig:tree_arch}, $p=8$, and the data flows from left to right. 
If the tree has more levels than hardware supports (three levels in the example; a problem size $2^{20}$ workload has 20 levels),
the remaining levels are scheduled and processed in DFS via the accumulator.
Outputs from the hardware tree (Level 3 in Figure~\ref{fig:tree_arch}) are pushed in an accumulator-local register file and once operands are ready, popped to the accumulator PE.
The bottom of Figure~\ref{fig:tree_arch} shows its schedule. 
Initially, there are gaps as the accumulator must wait two cycles for each input pair.
However, once multiple levels are processed, the gaps are filled;
this can be seen in cycle 44, where levels 4 and 5 are processed by the PE at the same time (i.e., both in the pipeline).
Thus the PEs in the tree have high utilization: for a $2^{20}$ workload, they are over 99\% utilized during the whole computation.
The red arrows coming out of each PE show how computations are sampled and output to create the Product MLE. 
The Forward Tree (Build MLE) support is shown in blue, and data flows from right to left.
Each PE takes the previous level and a challenge $r_i$ as inputs and generates two outputs that are fed to the next level.
Similarly, if the tree has more levels, the accumulator PE is scheduled to generate (roughly) one output per cycle to feed the rest of the PEs, corresponding to the last 3 levels, outputting 8 results in the end.
Switching is supported by muxes at PE inputs.

The advantages of this hybrid traversal \cite{mulfunc_tree} are noticeable:
it eliminates the need to store entire intermediate levels, making it practical for large problem sizes, and provides the ability to rate-match with upstream or downstream units and maintain throughput.
By adjusting the number of PEs, the unit can handle varying input and output rates, forming a full pipeline with other units.
The ability to reuse across multi-functions eliminates the need to allocate multiple dedicated units, saving 41.6\% area
across global Pareto design points in Section~\ref{sec:Evaluation}.
As HyperPlonk's code uses BFS, it experiences greater dependence distances. Our traversal mitigates this by scheduling work to avoid dependence stalls. Executing nodes already stresses CPU's limited compute resources, so we expect hybrid traversal to have little performance impact in software.

\subsection{Construct N\&D and FracMLE}
\label{sec:gen_nd}

\subsubsection{Creating $\phi$ with Constant-Time Inversion}
The Construct N\&D stage generates the $N$ and $D$ MLEs discussed in Section~\ref{sec:permcheck}.
Elements of six intermediate MLEs, $D_{1\dots3}$, and $N_{1\dots3}$, are computed in parallel from modular additions and multiplications of the witness ($w_{1\ldots3}$) and wiring permutation ($\sigma_{1\ldots3}$) MLEs stored in on-chip SRAM, and two challenges from SHA3.
These intermediate MLEs are written off-chip for the subsequent PermCheck and multiplied to obtain the $D$ and $N$ (e.g., $D[i] = D_1[i]D_2[i]D_3[i]$) elements, and fed to the FracMLE unit.

The Fraction MLE, $\phi$, requires computing the modular inverse of every element of the Denominator MLE ($D^{-1}$), and multiplying each inverted element with the corresponding element of the Numerator MLE ($N$).
Given $x$, modular inversion outputs a $y$, such that $x\cdot y \: \mathrm{mod}\: p = 1$.
We use the constant time \textit{Binary Extended Euclidean Algorithm (BEEA)}~\cite{Pornin_2020} for this operation to make it data-oblivious and ensure constant scheduling.
The algorithm requires $2W-1$ iterations of a loop consisting of shifts and subtractions, where $W$ is the number of input bits
($W=255$), resulting in a 509-cycle latency.
A data-dependent implementation is faster for smaller input values, but since we are computing on random inputs (derived from SHA3 hashes), the 
average latency of such an implementation 
(2$(\sum_{i=1}^{255}{\frac{255-i}{2^i}})-1 \approx 505$ cycles) is only 1$\%$ better. 
In exchange for this negligible latency overhead, our constant-time implementation reduces design complexity as this ensures elements of $D^{-1}$ (and thus $\phi$) are generated in-order if we execute multiple inverses in parallel (as we will discuss shortly).
Parallel execution of the data-dependent algorithm may output elements out-of-order and would require buffering or stalling to resolve.

\begin{figure}[t]
    \centering
    \includegraphics[width=\columnwidth]{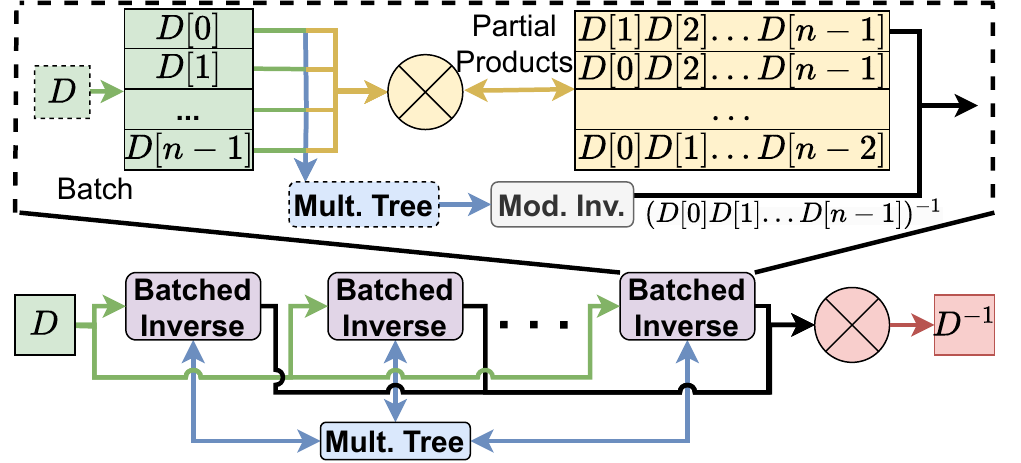}
    \caption{Modular inverse unit using multiple batched inverse units and a shared multiplier tree.}
    \vspace{-1.5em}
    \label{fig:frac_mle}
\end{figure}

\subsubsection{Batching for Modular Inversion}
Computing modular inverses for each 255-bit MLE element is expensive, so we heavily optimize it.
We leverage Montgomery batching~\cite{Montgomery_1987} to amortize the cost of one inversion across multiple elements and improve per-inversion throughput.
In the standard approach (also implemented in \cite{hyperplonk}), to compute the inverses of a batch of elements (e.g., $A, B, C, D$), partial products are computed sequentially (e.g., $A$, $AB$, $ABC$, $ABCD$) and the final product is inverted. 
This single inverse is then propagated backward to recover the individual inverses ($D^{-1}, C^{-1}$, etc.) through additional sequential multiplications. 
While this reduces the number of inversions, the sequential multiplications and high modular multiplication latency can limit hardware performance.

We address these limitations with two modifications.
First, we use a multiplier tree, detailed in Section~\ref{sec:multifunction_tree};
this significantly reduces latency and improves scaling for large values of the batch size, which we denote $b$, from $O(b)$ to $O(\log_2 b)$ multiplications.
Second, we overlap the sequential multiplications for partial products with modular inversion to mask long multiplication latencies.

\subsubsection{Batch Inversion Architecture}
\autoref{fig:frac_mle} illustrates the architecture of our modular inverse unit using batching.
We use multiple batched inversion units in round-robin fashion to completely mask long inversion latencies and enable the FracMLE unit to accept one input and generate one output per cycle, behaving as a pipeline with depth $b \times k$ where $k$ is the number of FracMLE units.
We achieve this by using enough batched inverse units to mask the latency of one batch inversion.
The multiplier tree and one multiplier are reused across all units to compute their inputs and produce the individual elements of $D^{-1}$, respectively.
One batch inversion latency is the maximum of the parallel partial product latency and the combined multiplier tree and modular inverse latencies.
The former scales $O(b)$ while the latter scales $O(\log_2 b)$.

\subsubsection{Optimizing for Batch Size}
\begin{figure}
    \centering
\includegraphics[width=\columnwidth]
    {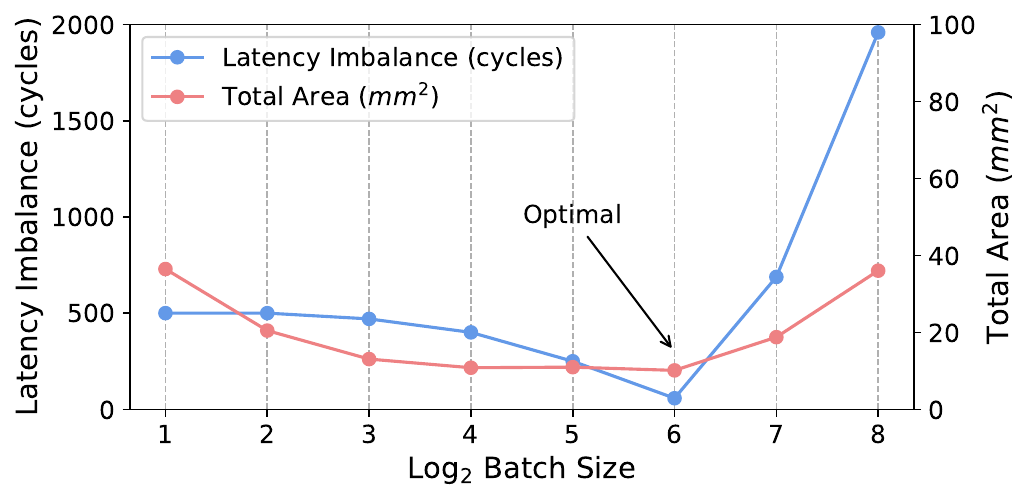}
    \vspace{-1em}
    \caption{Latency imbalance of batched inverse unit on the left (blue) and area cost of the FracMLE unit on the right (red). Both are optimal at 64. The area includes all hardware resources needed and does not account for area savings from reuse in the overall architecture.}
    \label{fig:batch_opt}
\end{figure}

We frame the choice of batch size $b$ as an optimization problem to minimize latency imbalance (between partial products and the multiplier tree and inversion) and area.
As shown in \autoref{fig:batch_opt}, the left y-axis plots latency imbalance, which is initially high due to the fixed cost of modular inversion. It decreases as partial product latency rises, reaching a minimum at $b = 64$, then increases again as partial-product latency begins to dominate. The right y-axis shows total area (including multiplier trees and partial products SRAM), which also reaches a minimum at $b = 64$ due to fewer required inverse units (e.g., 256 units for $b = 2$ vs. only 12 for $b = 64$).
Additionally, starting at $b = 64$, we can reuse the multiplier tree across all units since its $O(\log_2 b)$ latency allows it to complete one batch before the next arrives.
Beyond $b = 64$, partial-product latency surpasses modular inversion, and SRAM overheads continue to grow without reducing the number of inverse units—diminishing returns in both performance and area. Based on these tradeoffs, we select $b = 64$ as the optimal batch size.

\subsection{MLE Combine Unit}
The MLE Combine Unit is used in the Polynomial Opening step. As mentioned in Section \ref{sec:polyopen}, there are several linear combinations of MLEs that are performed before OpenCheck and before the MSMs. These operations are straightforward, and use a combination of MLEs stored in on-chip SRAM and off-chip memory to construct the MLEs used in subsequent steps. Because OpenCheck happens in series with the MSMs, the respective MLE Combine operations also happen in sequence. Consequently, we can share resources between these two operations. For the design point we highlight in Table \ref{tab:all_area_power}, without sharing, we would require 122 modular multipliers. With sharing, we require only 72, representing 41\% area savings.

\subsection{On-Chip MLE SRAM}
The MLE table size scales with the number of gates, i.e., for a problem size of $2^\mu$ gates, each MLE table has $2^\mu$ entries.
In practice, input MLEs are sparse. Control MLEs $q_L, q_R, q_M, q_O$ are all binary, and $q_C, w_1, w_2, w_3$ are roughly 90\% 1s and 0s and 10\% full bit-width.
These MLEs get reused throughout the protocol, so we store them on-chip in global SRAM.
We compress the tables, packing together control MLEs and using address translation units to perform lookups to either binary or 255-bit data.
These compression strategies save $10$ to $11\times$ on MLE storage across problem sizes.
This also significantly reduces off-chip accesses; for example, in the Polynomial Opening step, only 2 of 13 MLE tables ($\phi$ and $\pi$) are stored off-chip, cutting bandwidth usage by 84\%.

\section{zkSpeed Architecture $\&$ HyperPlonk Mapping}
\label{sec:overall_architecture}

\autoref{fig:dataflow_and_arch} provides an overview of the HyperPlonk protocol and zkSpeed architecture.
Colors indicate the mapping of protocol steps (Section~\ref{sec:HyperPlonk_Protocol_Steps}) to the accelerator units (Section~\ref{sec:HyperPlonk_Accelerator_Units}) they run on.
The zkSpeed architecture is streaming in nature and
captures on-chip data reuse when feasible via explicitly managed scratchpad memories.
The architecture has four major components: 
accelerator units,
local and global SRAM, a multi-channel shared bus, and HBM interface. 
HBM is needed to feed the chip with high bandwidths needed by HyperPlonk, and we conduct bandwidth sensitivity studies in Section~\ref{sec:Evaluation}.

\textbf{Dataflow.}
HyperPlonk is data oblivious at the stage granularity.
This allows zkSpeed to statically schedule computations and manage units, SRAM, and the buses via a simple controller.
Further, zkSpeed uses a shared bus rather than a crossbar or NoC.
This design choice was made after rigorously analyzing the HyperPlonk dataflow (see  Figure~\ref{fig:dataflow_and_arch}C).
We observed that at any given time, only 1-2 zkSpeed units typically communicate, and at most 4 independent bus channels are needed to avoid stalling
-- this is during Wire Identity where Construct N\&D sends results to FracMLE, FracMLE simultaneously feeds ProdMLE and MSM units, and ProdMLE streams to MSM.
Units overlap computation with each other, e.g., enabling MSM to start processing partial outputs from FracMLE, effectively masking latency.
Without bus stalls, we are able to rate match each accelerator unit to pipeline across modules when possible, further improving performance.

zkSpeed deploys a highly banked global SRAM and two local SRAMs for FracMLE and MSM units that store data unused by other units (the FracMLE SRAM captures MLE table reuse, the MSM's SRAM reuses elliptical curve points).
All other units share the global SRAM, which stores input MLEs.
At the start of execution, these MLEs are prefetched from HBM and remain unchanged on-chip throughout execution. They are read at the beginning of multiple protocol steps, thereby reducing HBM pressure.
zkSpeed allocates a single-channel shared bus for units to read the global SRAM since only one unit requires access at any given time.
HBM access is managed by a memory controller that has dedicated point-to-point connections to each module and the global SRAM, with enough wires to a given component to accommodate the widest access needed.
The controller interfaces with the HBM PHYs and arbitrates access to the HBM channels to ensure that no channel is being used more than once simultaneously (i.e., no channel conflict).

\textbf{zkSpeed Programmability.}
zkSpeed modules are programmed by instructions specifying problem sizes and configuring local controllers and address data from the buses, SRAM, and HBM.
Due to the ASIC nature, much of the fine-grained control is handled by FSMs within each unit.
For each HyperPlonk protocol step, each unit (e.g., bus, modules, global SRAM, and HBM controller) are configured with complex instructions and run to completion. Then next set of instructions are loaded to execute the next protocol step.

\textbf{HyperPlonk Trends and Outlook.}
Zero-Knowledge protocols are still continuously evolving and improving in several aspects. However, zkSpeed, with its focus on HyperPlonk, still has significant stability. Firstly, HyperPlonk has seen adoption both in industry implementations~\cite{hyperplonkespresso,binius} and academic research~\cite{zeromorph,binius,biniuslog}. 
In addition, our design is modular, and the key components of HyperPlonk SumChecks, MSMs, and MLEs, are present in essentially any modern SNARK protocol \cite{gkr_paper,spartan,groth,lasso,nova,protostar}. Therefore, zkSpeed can also be targeted to new protocols so long as they comprise the fabricated modules, including 
proof composition methods~\cite{ivc} that seek to compose protocols like Orion and HyperPlonk. 

\section{Methodology}

\subsection{Performance Modeling} SumCheck has a fixed, data-oblivious dataflow, which allows us to model its performance analytically. Modules that feed into the SumCheck units are also data-oblivious and are modeled similarly, accounting for rate-matching assumptions. 
For the MSM, we use a cycle-accurate simulator to model performance.
Each unit within zkSpeed is modularized. To understand full-chip performance, we conduct a design space exploration of all combinations of design parameters detailed in Table \ref{tab:design_space} and then analyze the pareto-optimal space to pick a suitable configuration for profiling runtimes. We also construct power traces to estimate average power for the full-chip architecture.
We use Catapult HLS 2023 to generate the RTL for Montgomery multipliers (as done in prior work~\cite{szkp}), the fully-pipelined, unified SumCheck unit (to handle ZeroCheck, PermCheck, and OpenCheck), and the fully-pipelined PADD unit. 
Consistent with HyperPlonk, we use the BLS12-381 elliptical curve, where all MLE datatypes (e.g., in the SumCheck unit) are 255-bit, and all elliptic curve points (e.g., in the PADD) are 381-bit.
Using Design Compiler with TSMC 22nm, we find the critical path in our design is the 381-bit PADD unit at 1.05ns.
We use Synopsys 22nm Memory Compiler to generate SRAM estimates.
For SHA3, we use the IP block from OpenCores \cite{opencores_sha3}.
We scale down to 7nm using scale factors of 3.6$\times$ for area, 3.3$\times$ for power, and 1.7$\times$ for delay (as in prior work~\cite{szkp}), and clock all zkSpeed accelerators at 1 GHz.

\subsection{Benchmarks}
\label{sec:workloads}
HyperPlonk was evaluated using \textit{mock circuit} workloads ~\cite{hyperplonk}, as there is no publicly available compiler to generate real workloads. 
We similarly use synthetically generated workloads to model the performance of our architecture, which is standard for ZKP benchmarks, as performance primarily depends on the size of the workload.
Similarly, GZKP \cite{gzkp} uses synthetic workloads to benchmark their implementation for workload sizes of $2^{22}$ and higher. 
NoCap~\cite{nocap} uses workloads from libsnark~\cite{libsnark} that prove relatively small circuits and scales them up to larger problem sizes because their proving times are dominated by fixed overheads on smaller ZKP circuits.
In the context of HyperPlonk, workload statistics primarily affect the witness commit (Sparse MSM) step. The scalar distributions of the MSMs in the Wiring Identity and Polynomial Opening are random because they are constructed in part from SHA3 challenges. Therefore, while all MSMs are data-dependent, the overall runtimes are effectively workload-agnostic and roughly the same at iso-problem size. All other steps in HyperPlonk are data-oblivious.
From prior work \cite{szkp, gzkp, pipezk, libsnark}, we know that in Sparse MSMs, the scalars are typically 5-10\% dense (i.e., full bit-width), and 95-99\% sparse.
Since the dense components of Sparse MSMs are still runtime dominant, we assume a pessimistic upper bound of 10\% dense scalars and 90\% sparse scalars, of which 45\% are 1s and 45\% are 0s. The rest of the protocol steps operate on 100\% dense scalars/MLE values.
In our evaluation, we use five workloads from prior work \cite{hyperplonk, libsnark}, and show our results in Table \ref{tab:benchmarks}.

\section{Evaluation}
\label{sec:Evaluation}

\subsection{Pareto Space Analysis}
\begin{figure}[t!]
\centerline{\includegraphics[width=\columnwidth]{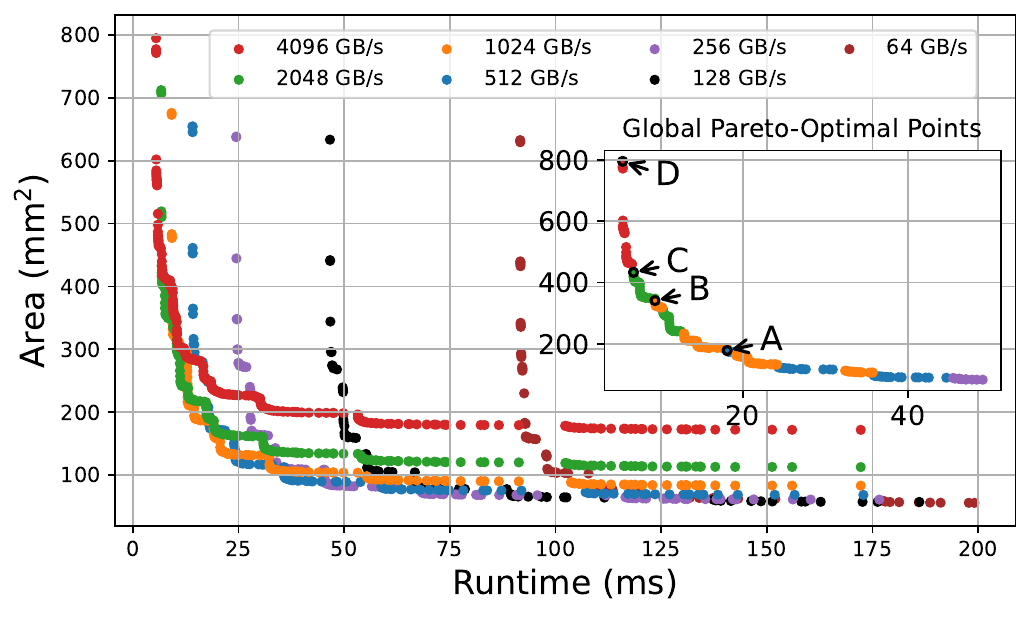}}
\vspace{-1em}
\caption{Pareto Frontiers for $2^{20}$ Gates. We plot the individual Pareto curves for each bandwidth model and the globally optimal Pareto curve (for designs under 50ms) in the inset.}
\Description{Pareto Frontiers for $2^{20}$ Gates. We plot both the individual Pareto curves for each bandwidth model, and then the globally optimal Pareto curve across all bandwidths in the inset.}
\vspace{-.5em}
\label{fig:Pareto_curve}
\end{figure}

Figure~\ref{fig:Pareto_curve} shows the design space for a problem size of $2^{20}$ gates under seven bandwidth scenarios. We sweep all the parameters in Table \ref{tab:design_space} and obtain Pareto curves for each bandwidth individually, and then construct the global Pareto curve from these seven local Pareto curves. 
The key highlight from this plot is that HBM3-scale bandwidths (e.g., 1-4 TB/s \cite{hbm3}) \textit{do} yield significant performance gains over, e.g., an HBM2-scale bandwidth (0.5 TB/s \cite{hbm2}).
Beyond 300 mm$^2$, the globally Pareto-optimal design configurations yield over 2$\times$ speedups compared to 512 GB/s designs and over 700$\times$ speedups over the CPU baseline. This is because high-performance SumCheck designs quickly saturate 512 GB/s of bandwidth. 
In this analysis, we also include the cost of HBM PHYs \cite{nocap, osiris, haac, hbm2, f1, clake, bts, ark, sharp} where the PHY cost is 14.9 mm$^2$ for a single HBM2 PHY and 29.6 mm$^2$ for a single HBM3 PHY.
For low-performance designs (below 100mm$^2$), high-bandwidth memory becomes less effective, as these designs tend to underutilize the available bandwidth while still incurring the high PHY area overhead associated with HBM-scale memory.
Figure~\ref{fig:Pareto_curve} shows that zkSpeed remains viable even at bandwidths typical of DDR5~\cite{micron_ddr5}  (256 GB/s and below).
While HBM allows exploring scalability benefits, less expensive memory technologies
can be used to achieve Pareto-optimality within a target performance range (e.g., within 50 ms).

\begin{figure}[t!]
\centerline{\includegraphics[width=\columnwidth]{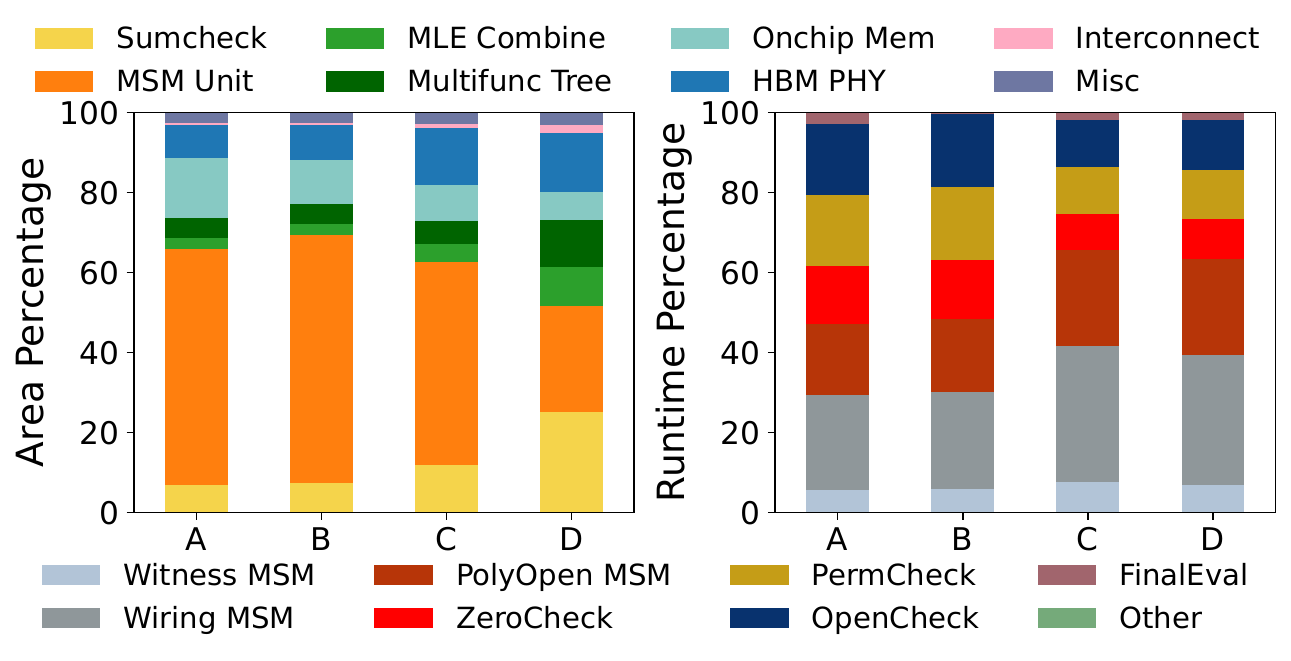}}
\vspace{-1em}
\caption{Area (left, top legend) and runtime (right, bottom legend) breakdown for the chosen Pareto points in \autoref{fig:Pareto_curve}.}
\Description{Area breakdown and runtime for the seven Pareto points we choose.}
\label{fig:area_breakdown}
\end{figure}

\begin{table}[t!]
\centering
\caption{Design Space of zkSpeed Architecture.}
\label{tab:design_space}
\footnotesize
\resizebox{\columnwidth}{!}{
\setlength{\tabcolsep}{1mm}{
\begin{tabular}{|c|c|c|}
\hline
\textbf{Module} & \textbf{Design Knob} & \textbf{Values}   \\ \hline \hline
\textbf{MSM} & Cores& 1, 2\\ \hline
\textbf{MSM} & PEs per Core& 1, 2, 4, 8, 16\\ \hline
\textbf{MSM} & Window Size   & 7, 8, 9, 10\\ \hline
\textbf{MSM} & Points/PE & 1K, 2K, 4K, 8K, 16K \\ \hline
\textbf{FracMLE} & PEs & 1, 2, 4 \\ \hline
\textbf{SumCheck} & PEs  & 1, 2, 4, 8, 16\\ \hline
\textbf{MLE Update} & PEs & 1, 2, $\ldots$ , 11 \\ \hline
\textbf{MLE Update} & Modmuls/PE & 1, 2, 4, 8, 16 \\ \hline
\textbf{Memory} & Bandwidth (GB/s) & 64, 128, 256, 512, 1T, 2T, 4T \\ \hline
\end{tabular}
}}
\vspace{-1em}
\end{table}

We analyze the area and runtime of selected points on the Pareto curve in \autoref{fig:area_breakdown} to further understand bandwidth sensitivity.
We pick Pareto points representing the highest-performing design point for each bandwidth level. 
In the area breakdown, moving from low to high-performance design points (A to D), the proportion of the SumCheck area increases significantly, 
because SumCheck is bandwidth-intensive, and higher bandwidth allows more parallel SumCheck PEs, boosting throughput. 
The MSM unit accounts for a large portion of the total area, but its absolute area remains unchanged when switching to high performance.
This trend is also evident in the runtime breakdown: total runtime decreases as bandwidth increases, and the runtime contributions of the SumCheck-related processes (ZeroCheck, PermCheck, and OpenCheck) become smaller.
Our analysis of the Pareto design points shows that 
high performance significantly depends on sufficient bandwidth, particularly improving the SumCheck computation. 
Conversely, for low-performance designs, the system utilizes less bandwidth and allocates more resources to MSM computation.

\subsection{Bandwidth Sensitivity}
\begin{figure}
  \centering
  \hspace{-5mm}
  \includegraphics[width=0.5\textwidth]{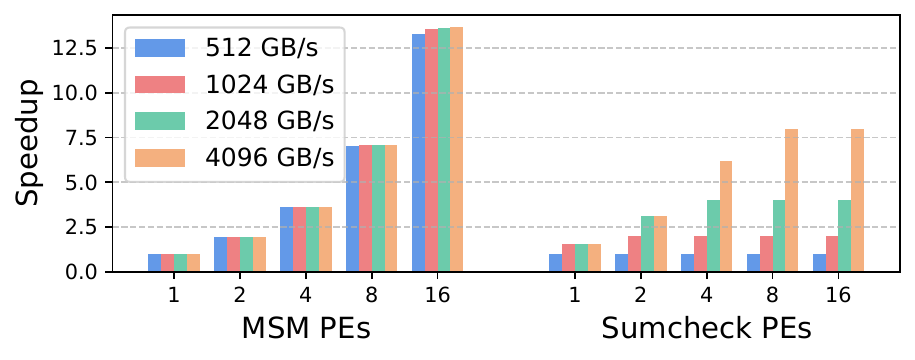}
  \vspace{-1.5em}
  \caption{Performance scaling with different memory technologies and PEs for MSM and SumCheck.
  Speedup normalized to 1 PE at 512 GB/s.}
  \label{fig:bw_sensitivity}
  \vspace{-1em}
\end{figure}

\autoref{fig:bw_sensitivity} shows how the speedups for MSM-related computation and SumCheck-related computation scale with increased PE count and bandwidth. These are two runtime-dominant components in our design points shown in \autoref{fig:area_breakdown}.
We take the runtime of all MSM and SumCheck operations for 1 PE under 512 GB/s, and compute respective speedups to these numbers.
Because MSMs are compute-bound, adding compute resources improves the speedups significantly, while adding bandwidth does not.
We do not see perfectly linear speedup because of the serialization incurred in Polynomial Opening MSMs. 
SumChecks, which rely on a streaming-based approach, are memory-bound. As we add compute, we see linear speedups initially and then diminishing returns after saturating bandwidth. Consequently, in our Pareto-optimal design space, we see most points along the curve outside high-performance regimes use at most 2 SumCheck PEs compared to 8 or 16 MSM PEs. 

\subsection{CPU Performance Comparisons}
Our CPU is an AMD EPYC 7502 32-core processor \cite{cpu_amd1, cpu_amd2, cpu_amd3, szkp}.
The total die size is 296 mm$^2$. We sweep the problem sizes, and for each problem size, pick a Pareto-optimal design point that is close to 296 mm$^2$.
In these comparisons, we exclude the PHY cost, since the AMD EPYC processor has its own separate die for I/O \cite{amd_article}.
Therefore, we compare our total compute and on-chip memory area with the CPU's total core area, including on-chip caches.
We assume 2 TB/s HBM to achieve Pareto-optimality in \autoref{fig:Pareto_curve}.

\subsubsection{Runtime Breakdown and Utilization}
\begin{figure}[t!]
\centerline{\includegraphics[width=\columnwidth]{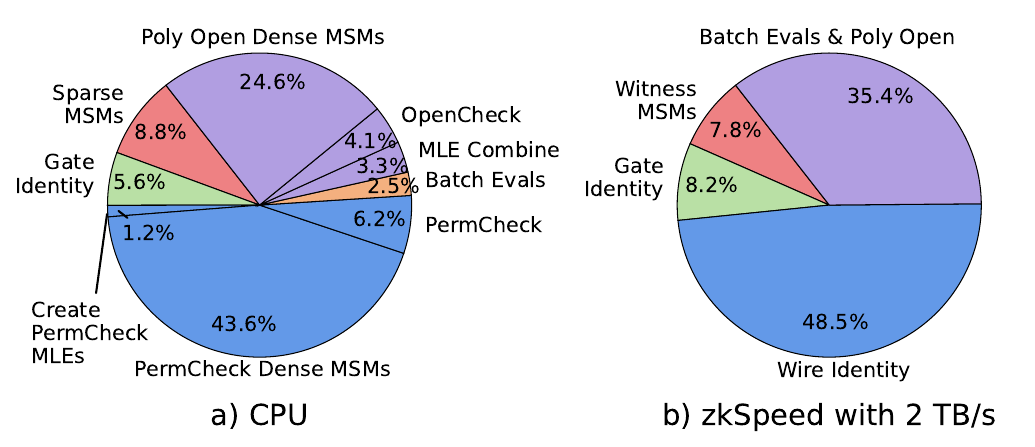}}
\vspace{-1em}
\caption{Runtime breakdown for CPU and zkSpeed at $2^{20}$ gates. CPU's sequential kernel execution enables finer breakdowns; aggregate step times are presented for zkSpeed.}
\Description{run time breakdown.}
\label{fig:runtime_breakdown_compare}
\end{figure}

\begin{figure}[t!]
\centerline{\includegraphics[width=\columnwidth]{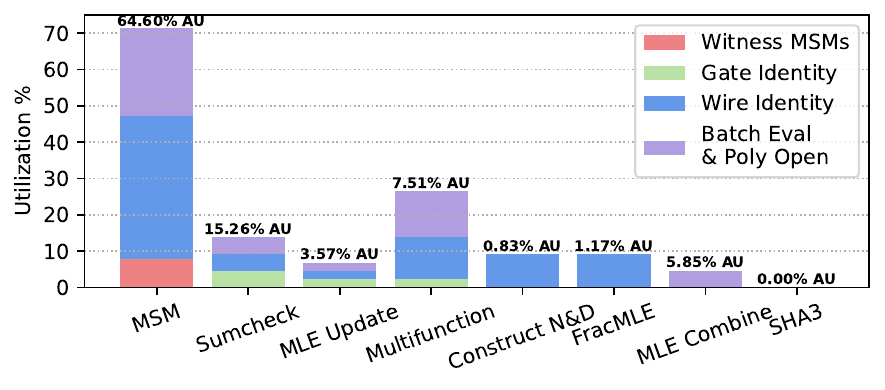}}
\vspace{-1.5em}
\caption{Utilization of zkSpeed modules, with compute area utilization (AU) listed on top of each bar. 
Stacked bar colors reflect the protocol steps in which each module is active.}
\Description{utilization breakdown.}
\vspace{-1em}
\label{fig:hardware_utilization_breakdown}
\end{figure}

Figure~\ref{fig:runtime_breakdown_compare} shows the latency breakdown of HyperPlonk proving on a CPU and a pareto-optimal zkSpeed design for a problem size of $2^{20}$ gates.
The CPU executes kernels sequentially, enabling detailed profiling; we report step latency for zkSpeed due to its parallel scheduling of kernels.
As expected, the majority of time goes to processing MSMs, while a handful of other kernels account for single percentage points of runtime.
Figure ~\ref{fig:hardware_utilization_breakdown} presents the utilization of each unit and relative (datapath) area allocation (design in Table~\ref{tab:all_area_power}).
The utilizations vary from over 70\% to 5\% for some modules.
zkSpeed was intentionally designed (via the design space search in Figure~\ref{fig:Pareto_curve}) to allocate resources to cores to optimize high performance per area, i.e. the Pareto front.
This can be seen in our analysis in two ways.
First, the cores taking up most area, notably MSM at 64.6\%, are the most used, and following the profiling data (\autoref{fig:runtime_breakdown_compare}) require the most speedup.
Second, though some units are used infrequently, they (i) take up little area and (ii) are essential to accelerate to achieve the speedups desired (i.e., 2-3 orders of magnitude).
For example, the SHA-3 unit is rarely used, but provides a speedup of over $300\times$ over the CPU and takes only $ 5888 \mu m^2$ area.
Additionally, consider that MLE Combine makes up 3.3\% of the CPU runtime, but without acceleration caps speedup to a mere 30.3$\times$, thus justifying its 5.85\% area allocation and relatively low utilization.

\subsubsection{Iso-CPU Area Comparisons}
\begin{figure}
  \centering
  \hspace{-5mm}
  \includegraphics[width=0.5\textwidth]{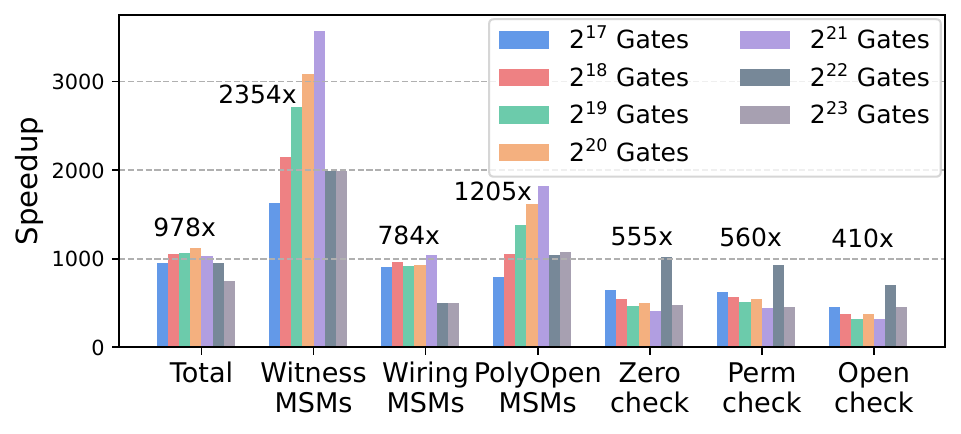}
  \vspace{-1.5em}
  \caption{Speedup over CPU at Iso-CPU Area Designs. Each problem size has a different Pareto-optimal point. Each bar reflects absolute speedup, while the annotated speedup is the gmean computed across gate counts for each kernel.}
  \label{fig:iso_cpu_speedup}
\end{figure}

After picking each Pareto-optimal design, we run synthetic benchmarks over problem sizes $2^{17}-2^{23}$. \autoref{fig:iso_cpu_speedup} shows the speedup of each design over CPU baseline, and the breakdown across different steps of the protocol to understand where our speedups come from. 
In general, we get more speedup from our MSM units than the SumCheck units.
This intuitively follows, given our observations that MSMs are compute-bound and more robust to bandwidth constraints.
Additionally, the CPU poorly handles sparse computations because it serially computes the point addition for $1$-valued scalars. Polynomial Opening MSMs also incur serialization costs that we reduce by overlapping MSM executions where possible. 
The variations in speedups over different problem sizes are an artifact of our choice to highlight different Pareto points per problem size; for example, at $2^{20}$ size problems, a dual-core MSM is used, while at $2^{22}$ size a single-core MSM is chosen.
This is because on-chip MLE SRAM area begins to dominate, limiting MSM compute area and therefore achievable speedup. Storing MLE tables entirely off-chip may improve  MSM speedups at higher SumCheck bandwidth costs. These tradeoffs can be explored in future work.

\begin{table}[]
\centering
\caption{zkSpeed evaluation on real-world workloads.}
\vspace{-1em}
\label{tab:benchmarks}
\resizebox{1\columnwidth}{!}{
\setlength{\tabcolsep}{1mm}{
\begin{tabular}{|c|c|cc|}
\hline
\multirow{2}{*}{\textbf{Workload}} & \multirow{2}{*}{\textbf{\begin{tabular}[c]{@{}c@{}}Problem\\ Size\end{tabular}}} & \multicolumn{2}{c|}{\textbf{Runtime (ms)}}\\ \cline{3-4} 
   &  & \multicolumn{1}{c|}{\textbf{CPU}} & \textbf{zkSpeed}\\ \hline
Zcash  & $2^{17}$ & \multicolumn{1}{c|}{1429}   & 1.984  (\textbf{720$\times$}) \\ \hline
Auction& $2^{20}$ & \multicolumn{1}{c|}{8619}   & 11.405 (\textbf{755$\times$})  \\ \hline
$2^{12}$ Rescue-Hash Invocations   & $2^{21}$ & \multicolumn{1}{c|}{18637}  & 22.082 (\textbf{844$\times$})  \\ \hline
Zexe's Recursive Circuit   & $2^{22}$ & \multicolumn{1}{c|}{37469}  & 43.451 (\textbf{862$\times$})  \\ \hline
Rollup of 10 Pvt Tx& $2^{23}$ & \multicolumn{1}{c|}{74052}  & 86.181 (\textbf{859$\times$})  \\ \hline
\end{tabular}
}
}
\end{table}

\begin{table}[]
\centering
\caption{Comparison of zkSpeed with Prior ZKP Accelerators on $2^{24}$ Constraints/Gates. N = NTT, S = SumCheck, M = MSM}
\vspace{-1em}
\label{tab:megatable}
\resizebox{1\columnwidth}{!} {
\begin{tabular}{@{}cccc@{}}
\toprule
\textbf{Accelerator} & NoCap & SZKP+ & zkSpeed \\ \midrule
\textbf{Protocol} & Spartan+Orion & Groth16 & HyperPlonk \\ \hline
\textbf{Main Kernels} & N \& S & N \& M & S \& M \\ \hline
\textbf{Encoding} & R1CS & R1CS & Plonk \\ \hline
\textbf{Proof Size} & 8.1 MB & 0.18 KB & 5.09 KB \\ \hline
\textbf{Setup} & none & circuit-specific & universal \\ \hline
\textbf{Prime} & fixed & arbitrary & arbitrary \\ \hline
\textbf{Bit-width} & 64 & 255b/381b & 255b/381b \\ \hline
\textbf{CPU Prover (s)} & 94.2 & 51.18 & 145.5 \\ \hline
\textbf{HW Prover (ms)} & 151.3 & 28.43 & 171.61 \\ \hline
\textbf{Verifier (ms)} & 134 & 4.2 & 26 \\ \hline
\textbf{Chip Area (mm$^2$)} & 38.73 & 353.2 & 366.46 \\ \hline
\textbf{\# Modmuls} & 2432 & 1720 & 1206 \\ \hline
\textbf{Modmul (mm$^2$)} & 0.002 & 0.133 / 0.314 & 0.133 / 0.314 \\ \hline
\textbf{Power (W)} & 62 & >220 W & 170.88 \\ 
\bottomrule
\end{tabular}
}
\end{table}

\subsection{Workload Evaluation}\label{work_eval}
We pick a fixed design and show the end-to-end speedups in Table \ref{tab:benchmarks}. As mentioned in Section \ref{sec:workloads}, we assume pessimistic 10\% sparse scalar statistics for each workload. Our  design has one MSM unit with 9-bit windows, 16 PEs, and 2048 points per PE, with 1 FracMLE PE, 2 SumCheck PEs, 11 MLE Update PEs, and 4 modular multipliers per MLE Update PE. The area breakdown is provided in Table \ref{tab:all_area_power}. 
At roughly iso-CPU-core compute cost, zkSpeed achieves a geometric mean speedup of 801$\times$ over the CPU, with total area of 366.46 mm$^2$, total average power of 170.88 W, and total power density of 0.46 W/mm$^2$, which is within that of our CPU \cite{szkp}.

%%%%%%%%%%%%%%%%%%%%%%%%%%%%%%%%%%%%%%%%%%%%%%%%%%%%%%%%%
\begin{table}[]
\caption{Area and power of zkSpeed. 
Other includes the SHA3 unit and interconnect.}
\vspace{-1em}
\label{tab:all_area_power}
% \footnotesize
\resizebox{0.96\columnwidth}{!}{
\setlength{\tabcolsep}{1mm}{
\begin{tabular}{lrr}
\toprule
\textbf{Module} & \multicolumn{1}{l}{\textbf{Area (mm²)}} & \multicolumn{1}{l}{\textbf{Average Power (W)}} \\ \midrule
MSM (16 PEs)   & 105.64   & 76.19 \\
SumCheck (2 PEs)   & 24.96& 5.38  \\
Construct N\&D & 1.35 & 0.19   \\
FracMLE   & 1.92 & 0.25   \\
MLE Combine& 9.56 & 0.34  \\
MLE Update & 5.84 & 1.13   \\
Multifunction Tree & 12.28& 4.16  \\
Other  & 1.98 & 0.04   \\ \hline
\textbf{Total Compute} & \textbf{163.53}  & \textbf{87.68} \\ \hline
SRAM   & 143.73   & 19.60  \\
HBM3 (2 PHYs)  & 59.20& 63.60  \\
\textbf{Total Memory}  & \textbf{202.93}  & \textbf{83.20} \\ \hline
\textbf{Total} & \textbf{366.46}  & \textbf{170.88} \\ \bottomrule
\end{tabular}
}
}
\vspace{-1em}
\end{table}
%%%%%%%%%%%%%%%%%%%%%%%%%%%%%%%%%%%%%%%%%%%%%%%%%%%%%%%%%
\section{Related Work}
\label{sec:related_work}
Much of the prior body of cryptographic hardware and systems research has focused on Fully Homomorphic Encryption and Multi-Party Computation \cite{bts, ark, sharp, f1, clake, rpu, haac, karthik, ciflow, orion_fhe, jha2024deepreshape}.
ZKP hardware research is relatively newer, and has focused primarily on accelerating NTTs and MSMs~\cite{priorMSM, distMSM, cuZK, gypso, reZK, myotosis, MSMAC, intel_zkp, elastic_msm, tches_ntt_msm, sam, legozk, unizk, graz, zhang2025code}.
A few works have accelerated SumChecks on GPU \cite{batchzk} and ASIC \cite{nocap} as well as hashing alternatives to SHA-based hash functions \cite{gottahashemall, amaze, unizk}.
Some systems accelerate end-to-end Groth16 proofs (using NTTs and MSMs) on GPU~\cite{gzkp} and ASICs \cite{szkp, pipezk}, while others accelerate Spartan proofs with Orion commitments (using NTTs and SumChecks) \cite{nocap}. 
We compare zkSpeed with two ASICs, SZKP and NoCap, that accelerate full proofs end-to-end.

\textbf{SZKP} is the state-of-the-art for accelerating Groth16 proofs, focusing on scalable MSM designs and (quasi)-deterministic scheduling for Pippenger's algorithm. It accelerates all MSMs, including Sparse G2 MSMs (not present in HyperPlonk), achieving geomean speedups of 493$\times$ over a CPU. 
SZKP improves on PipeZK~\cite{pipezk}, the first hardware accelerator for Groth16 proofs.
While Groth16 and HyperPlonk have similar application spaces, as mentioned in Section \ref{sec:intro}, the key advantage of using HyperPlonk is the universal setup, which means that the protocol parameters are application-agnostic. For Groth16, \textit{every new application} that wants to use a ZKP needs its own trusted setup ceremony \cite{ceremony}, which is impractical as the application space grows.
Given this context and the recent shift away from Groth16 \cite{trusted_set_up}, the slightly larger proof sizes are considered a reasonable tradeoff. 

\textbf{NoCap} is a vector-based processor for accelerating Spartan+Orion proofs, but its application space differs from zkSpeed's.
NoCap thrives in applications where proof size is not critical, or there are few verifiers.
It achieves $41\times$ geomean speedups over PipeZK. In contrast, zkSpeed is ideal for many verifiers and in consensus-based systems; this is where ZKPs are experiencing growing interest.

\textbf{Comparison}: \autoref{tab:megatable} compares zkSpeed, NoCap, and SZKP's protocols, software, and hardware costs. zkSpeed's parent HyperPlonk has the slowest software prover, reflecting the complexity of the protocol. Of note, Spartan's prover is slow; NoCap's authors explain this is due to inefficient implementation.

We compare NoCap's hardware implementation using the design point and numbers from their paper scaled to 7nm using scale factors from prior work \cite{haac, szkp}. 
We then select a zkSpeed configuration with roughly similar prover time to NoCap. At iso-prover time, zkSpeed incurs a nearly $10\times$ area cost in return for a three orders-of-magnitude reduction in proof size. 
NoCap's lower costs come from not having MSMs, having simpler SumChecks, and using
a 64-bit Goldilocks-64 prime field that yields smaller modmuls. 
Consequently, NoCap runs all operations several times, including SumChecks 3 times, to obtain 128 bits of security.
In contrast, zkSpeed supports arbitrary 255-bit and 381-bit primes for MLEs and elliptic curves points, respectively.
We further compare zkSpeed with an iso-area SZKP (Groth16) implementation, giving them the benefit of zkSpeed's improved MSMs, and optimistically scale up their design to use the BLS12-381 curve. 
This design, SZKP+, enjoys a 6$\times$ reduction in proving time compared to zkSpeed, largely because it has fewer MSMs on its critical path.
These speedups come at the cost of circuit-specific setup, incurring large costs any time the application is updated.
In sum, NoCap, SZKP, and zkSpeed address different application domains, representing a range of tradeoffs ranging from security properties to software/hardware costs.  

\textbf{Jellyfish}: Jellyfish is a HyperPlonk variant supporting gates of arity (fan-in) higher than 2. Unlike R1CS, it supports higher degree constraints, e.g. $x^7=y^5+y^2+7$. The additional expressiveness means, iso-application, the total size of all MLE tables decreases (the number of tables increases with arity, but table size decreases super-proportionally). High-degree gates have utility in many applications \cite{garuda}; this is especially pronounced when proving the correctness of cryptographic operations like encryption \cite{verizexe} or hash-functions \cite{poseidon}. zkSpeed could be extended to support Jellyfish, in which case the ratio of table count to table size may improve the runtime (with sufficient bandwidth). We leave this for future work.

\section{Conclusion}
This paper presents zkSpeed, the \textit{first} work to accelerate HyperPlonk proofs in hardware, which offers $O(n)$ time complexity compared to prominent zkSNARKs that rely on computational primitives that have $O(n \log n)$ complexity (e.g. Groth16).
zkSpeed constitutes accelerator units for all core HyperPlonk functions, with special attention paid to prominent kernels: SumCheck and MSM.
zkSpeed is a modular architecture, and we leverage a performance model to conduct design space optimization, and analyze the pareto frontier to identify well performing designs.
A zkSpeed accelerator with 366 mm$^2$ and 2 TB/s of bandwidth achieves geomean speedup of $801\times$ over CPU baselines, demonstrating the promise zkSpeed offers to accelerate HyperPlonk.


\begin{acks}
This research was developed with funding in part from the NSF CAREER award \#2340137, an NSF RINGS Award, from DARPA under the Data Protection in Virtual Environments (DPRIVE) program, grant number HR0011-21-9-0003, and the Hybrid Electromagnetic side-channel and Interactive-proof Methods to Detect and Amend LogicaL Rifts (HEIMDALLR) programs, grant number HR0011-25-C-0300, and from a gift award from Google.
\end{acks}


\bibliographystyle{ACM-Reference-Format}
\bibliography{Reference-base}


\begin{thebibliography}{76}

%%% the Web address.

\ifx \showCODEN    \undefined \def \showCODEN     #1{\unskip}     \fi
\ifx \showDOI      \undefined \def \showDOI       #1{#1}\fi
\ifx \showISBNx    \undefined \def \showISBNx     #1{\unskip}     \fi
\ifx \showISBNxiii \undefined \def \showISBNxiii  #1{\unskip}     \fi
\ifx \showISSN     \undefined \def \showISSN      #1{\unskip}     \fi
\ifx \showLCCN     \undefined \def \showLCCN      #1{\unskip}     \fi
\ifx \shownote     \undefined \def \shownote      #1{#1}          \fi
\ifx \showarticletitle \undefined \def \showarticletitle #1{#1}   \fi
\ifx \showURL      \undefined \def \showURL       {\relax}        \fi
% invisible to TeX
\providecommand\bibfield[2]{#2}
\providecommand\bibinfo[2]{#2}
\providecommand\natexlab[1]{#1}
\providecommand\showeprint[2][]{arXiv:#2}

\bibitem[lib(2018)]%
        {libsnark}
 \bibinfo{year}{2018}\natexlab{}.
\newblock \showarticletitle{libsnark: a C++ library for zkSNARK proofs.}
\newblock
\urldef\tempurl%
\url{https://github.com/scipr-lab/libsnark}
\showURL{%
\tempurl}


\bibitem[hbm(2023)]%
        {hbm3}
 \bibinfo{year}{2023}\natexlab{}.
\newblock \bibinfo{booktitle}{\emph{High Bandwidth Memory DRAM (HBM3)}}.
\newblock \bibinfo{type}{Technical Report} JESD238A. \bibinfo{institution}{JEDEC}.
\newblock
\newblock
\shownote{\url{https://www.jedec.org/standards-documents/docs/jesd238a}}.


\bibitem[cer(2023)]%
        {ceremony}
 \bibinfo{year}{2023}\natexlab{}.
\newblock \bibinfo{title}{What is the ZCash Ceremony? The Complete Beginners Guide}.
\newblock \bibinfo{howpublished}{\url{https://coinbureau.com/education/zcash-ceremony/}}.
\newblock


\bibitem[Abbaszadeh et~al\mbox{.}(2024)]%
        {10.1145/3658644.3670316}
\bibfield{author}{\bibinfo{person}{Kasra Abbaszadeh}, \bibinfo{person}{Christodoulos Pappas}, \bibinfo{person}{Jonathan Katz}, {and} \bibinfo{person}{Dimitrios Papadopoulos}.} \bibinfo{year}{2024}\natexlab{}.
\newblock \showarticletitle{Zero-Knowledge Proofs of Training for Deep Neural Networks}. In \bibinfo{booktitle}{\emph{Proceedings of the 2024 on ACM SIGSAC Conference on Computer and Communications Security}} (Salt Lake City, UT, USA) \emph{(\bibinfo{series}{CCS '24})}. \bibinfo{publisher}{Association for Computing Machinery}, \bibinfo{address}{New York, NY, USA}, \bibinfo{pages}{4316–4330}.
\newblock
\showISBNx{9798400706363}
\urldef\tempurl%
\url{https://doi.org/10.1145/3658644.3670316}
\showDOI{\tempurl}


\bibitem[Ahmed et~al\mbox{.}(2024)]%
        {amaze}
\bibfield{author}{\bibinfo{person}{Anees Ahmed}, \bibinfo{person}{Nojan Sheybani}, \bibinfo{person}{Davi Moreno}, \bibinfo{person}{Nges~Brian Njungle}, \bibinfo{person}{Tengkai Gong}, \bibinfo{person}{Michel Kinsy}, {and} \bibinfo{person}{Farinaz Koushanfar}.} \bibinfo{year}{2024}\natexlab{}.
\newblock \bibinfo{title}{AMAZE: Accelerated MiMC Hardware Architecture for Zero-Knowledge Applications on the Edge}.
\newblock
\newblock
\showeprint[arxiv]{2411.06350}~[cs.CR]
\urldef\tempurl%
\url{https://arxiv.org/abs/2411.06350}
\showURL{%
\tempurl}
\newblock
\shownote{Accepted to ICCAD 2024}.


\bibitem[AMD(2024)]%
        {cpu_amd2}
\bibfield{author}{\bibinfo{person}{Product AMD}.} \bibinfo{year}{2024}\natexlab{}.
\newblock \bibinfo{title}{Server Processor Specifications}.
\newblock
\newblock
\urldef\tempurl%
\url{https://www.amd.com/en/products/specifications/server-processor.html}
\showURL{%
\tempurl}


\bibitem[Ben{-}Sasson et~al\mbox{.}(2014)]%
        {zerocash}
\bibfield{author}{\bibinfo{person}{Eli Ben{-}Sasson}, \bibinfo{person}{Alessandro Chiesa}, \bibinfo{person}{Christina Garman}, \bibinfo{person}{Matthew Green}, \bibinfo{person}{Ian Miers}, \bibinfo{person}{Eran Tromer}, {and} \bibinfo{person}{Madars Virza}.} \bibinfo{year}{2014}\natexlab{}.
\newblock \showarticletitle{Zerocash: Decentralized Anonymous Payments from Bitcoin}. In \bibinfo{booktitle}{\emph{2014 {IEEE} Symposium on Security and Privacy, {SP} 2014, Berkeley, CA, USA, May 18-21, 2014}}. \bibinfo{publisher}{{IEEE} Computer Society}, \bibinfo{pages}{459--474}.
\newblock
\urldef\tempurl%
\url{https://doi.org/10.1109/SP.2014.36}
\showDOI{\tempurl}


\bibitem[Berrut and Trefethen(2004)]%
        {barycentric}
\bibfield{author}{\bibinfo{person}{Jean-Paul Berrut} {and} \bibinfo{person}{Lloyd~N Trefethen}.} \bibinfo{year}{2004}\natexlab{}.
\newblock \showarticletitle{Barycentric lagrange interpolation}.
\newblock \bibinfo{journal}{\emph{SIAM review}} \bibinfo{volume}{46}, \bibinfo{number}{3} (\bibinfo{year}{2004}), \bibinfo{pages}{501--517}.
\newblock


\bibitem[B{\"{u}}nz and Chen(2023)]%
        {protostar}
\bibfield{author}{\bibinfo{person}{Benedikt B{\"{u}}nz} {and} \bibinfo{person}{Binyi Chen}.} \bibinfo{year}{2023}\natexlab{}.
\newblock \showarticletitle{Protostar: Generic Efficient Accumulation/Folding for Special-Sound Protocols}. In \bibinfo{booktitle}{\emph{Advances in Cryptology - {ASIACRYPT} 2023 - 29th International Conference on the Theory and Application of Cryptology and Information Security, Guangzhou, China, December 4-8, 2023, Proceedings, Part {II}}} \emph{(\bibinfo{series}{Lecture Notes in Computer Science}, Vol.~\bibinfo{volume}{14439})}, \bibfield{editor}{\bibinfo{person}{Jian Guo} {and} \bibinfo{person}{Ron Steinfeld}} (Eds.). \bibinfo{publisher}{Springer}, \bibinfo{pages}{77--110}.
\newblock
\urldef\tempurl%
\url{https://doi.org/10.1007/978-981-99-8724-5\_3}
\showDOI{\tempurl}


\bibitem[Butt et~al\mbox{.}(2024)]%
        {intel_zkp}
\bibfield{author}{\bibinfo{person}{Shahzad~Ahmad Butt}, \bibinfo{person}{Benjamin Reynolds}, \bibinfo{person}{Veeraraghavan Ramamurthy}, \bibinfo{person}{Xiao Xiao}, \bibinfo{person}{Pohrong Chu}, \bibinfo{person}{Setareh Sharifian}, \bibinfo{person}{Sergey Gribok}, {and} \bibinfo{person}{Bogdan Pasca}.} \bibinfo{year}{2024}\natexlab{}.
\newblock \bibinfo{title}{if-ZKP: Intel FPGA-Based Acceleration of Zero Knowledge Proofs}.
\newblock
\newblock
\showeprint[arxiv]{2412.12481}~[cs.AR]
\urldef\tempurl%
\url{https://arxiv.org/abs/2412.12481}
\showURL{%
\tempurl}


\bibitem[Chen et~al\mbox{.}(2022)]%
        {hyperplonk}
\bibfield{author}{\bibinfo{person}{Binyi Chen}, \bibinfo{person}{Benedikt Bünz}, \bibinfo{person}{Dan Boneh}, {and} \bibinfo{person}{Zhenfei Zhang}.} \bibinfo{year}{2022}\natexlab{}.
\newblock \bibinfo{title}{{HyperPlonk}: Plonk with Linear-Time Prover and High-Degree Custom Gates}.
\newblock \bibinfo{howpublished}{Cryptology {ePrint} Archive, Paper 2022/1355}.
\newblock
\urldef\tempurl%
\url{https://eprint.iacr.org/2022/1355}
\showURL{%
\tempurl}


\bibitem[Chen et~al\mbox{.}(2024)]%
        {tches_ntt_msm}
\bibfield{author}{\bibinfo{person}{Xiangren Chen}, \bibinfo{person}{Bohan Yang}, \bibinfo{person}{Wenping Zhu}, \bibinfo{person}{Hanning Wang}, \bibinfo{person}{Qichao Tao}, \bibinfo{person}{Shuying Yin}, \bibinfo{person}{Min Zhu}, \bibinfo{person}{Shaojun Wei}, {and} \bibinfo{person}{Leibo Liu}.} \bibinfo{year}{2024}\natexlab{}.
\newblock \showarticletitle{A High-performance NTT/MSM Accelerator for Zero-knowledge Proof Using Load-balanced Fully-pipelined Montgomery Multiplier}.
\newblock \bibinfo{journal}{\emph{IACR Transactions on Cryptographic Hardware and Embedded Systems}} \bibinfo{volume}{2025}, \bibinfo{number}{1} (\bibinfo{date}{Dec.} \bibinfo{year}{2024}), \bibinfo{pages}{275–313}.
\newblock
\urldef\tempurl%
\url{https://doi.org/10.46586/tches.v2025.i1.275-313}
\showDOI{\tempurl}


\bibitem[Chillotti et~al\mbox{.}(2020)]%
        {chillotti2020tfhe}
\bibfield{author}{\bibinfo{person}{Ilaria Chillotti}, \bibinfo{person}{Nicolas Gama}, \bibinfo{person}{Mariya Georgieva}, {and} \bibinfo{person}{Malika Izabach{\`e}ne}.} \bibinfo{year}{2020}\natexlab{}.
\newblock \showarticletitle{TFHE: fast fully homomorphic encryption over the torus}.
\newblock \bibinfo{journal}{\emph{Journal of Cryptology}} \bibinfo{volume}{33}, \bibinfo{number}{1} (\bibinfo{year}{2020}), \bibinfo{pages}{34--91}.
\newblock


\bibitem[Daftardar et~al\mbox{.}(2024)]%
        {szkp}
\bibfield{author}{\bibinfo{person}{Alhad Daftardar}, \bibinfo{person}{Brandon Reagen}, {and} \bibinfo{person}{Siddharth Garg}.} \bibinfo{year}{2024}\natexlab{}.
\newblock \showarticletitle{SZKP: A Scalable Accelerator Architecture for Zero-Knowledge Proofs}. In \bibinfo{booktitle}{\emph{Proceedings of the 2024 International Conference on Parallel Architectures and Compilation Techniques}}. \bibinfo{pages}{271--283}.
\newblock


\bibitem[de~Castro et~al\mbox{.}(2021)]%
        {does_fhe_need_accelerators}
\bibfield{author}{\bibinfo{person}{Leo de Castro}, \bibinfo{person}{Rashmi Agrawal}, \bibinfo{person}{Rabia Yazicigil}, \bibinfo{person}{Anantha Chandrakasan}, \bibinfo{person}{Vinod Vaikuntanathan}, \bibinfo{person}{Chiraag Juvekar}, {and} \bibinfo{person}{Ajay Joshi}.} \bibinfo{year}{2021}\natexlab{}.
\newblock \bibinfo{title}{Does Fully Homomorphic Encryption Need Compute Acceleration?}
\newblock \bibinfo{howpublished}{Cryptology {ePrint} Archive, Paper 2021/1636}.
\newblock
\urldef\tempurl%
\url{https://eprint.iacr.org/2021/1636}
\showURL{%
\tempurl}


\bibitem[Dellepere et~al\mbox{.}(2024)]%
        {garuda}
\bibfield{author}{\bibinfo{person}{Michel Dellepere}, \bibinfo{person}{Pratyush Mishra}, {and} \bibinfo{person}{Alireza Shirzad}.} \bibinfo{year}{2024}\natexlab{}.
\newblock \bibinfo{title}{Garuda and Pari: Faster and Smaller {SNARKs} via Equifficient Polynomial Commitments}.
\newblock \bibinfo{howpublished}{Cryptology {ePrint} Archive, Paper 2024/1245}.
\newblock
\urldef\tempurl%
\url{https://eprint.iacr.org/2024/1245}
\showURL{%
\tempurl}


\bibitem[Diamond and Posen(2023)]%
        {binius}
\bibfield{author}{\bibinfo{person}{Benjamin~E. Diamond} {and} \bibinfo{person}{Jim Posen}.} \bibinfo{year}{2023}\natexlab{}.
\newblock \bibinfo{title}{Succinct Arguments over Towers of Binary Fields}.
\newblock \bibinfo{howpublished}{Cryptology {ePrint} Archive, Paper 2023/1784}.
\newblock
\urldef\tempurl%
\url{https://eprint.iacr.org/2023/1784}
\showURL{%
\tempurl}


\bibitem[Diamond and Posen(2024)]%
        {biniuslog}
\bibfield{author}{\bibinfo{person}{Benjamin~E. Diamond} {and} \bibinfo{person}{Jim Posen}.} \bibinfo{year}{2024}\natexlab{}.
\newblock \showarticletitle{Polylogarithmic Proofs for Multilinears over Binary Towers}.
\newblock \bibinfo{journal}{\emph{{IACR} Cryptol. ePrint Arch.}} (\bibinfo{year}{2024}), \bibinfo{pages}{504}.
\newblock
\urldef\tempurl%
\url{https://eprint.iacr.org/2024/504}
\showURL{%
\tempurl}


\bibitem[Ebel et~al\mbox{.}(2025)]%
        {orion_fhe}
\bibfield{author}{\bibinfo{person}{Austin Ebel}, \bibinfo{person}{Karthik Garimella}, {and} \bibinfo{person}{Brandon Reagen}.} \bibinfo{year}{2025}\natexlab{}.
\newblock \showarticletitle{Orion: A Fully Homomorphic Encryption Framework for Deep Learning}. In \bibinfo{booktitle}{\emph{Proceedings of the 30th ACM International Conference on Architectural Support for Programming Languages and Operating Systems, Volume 2}} (Rotterdam, Netherlands) \emph{(\bibinfo{series}{ASPLOS '25})}. \bibinfo{publisher}{Association for Computing Machinery}, \bibinfo{address}{New York, NY, USA}, \bibinfo{pages}{734–749}.
\newblock
\showISBNx{9798400710797}
\urldef\tempurl%
\url{https://doi.org/10.1145/3676641.3716008}
\showDOI{\tempurl}


\bibitem[Ebel and Reagen(2024)]%
        {osiris}
\bibfield{author}{\bibinfo{person}{Austin Ebel} {and} \bibinfo{person}{Brandon Reagen}.} \bibinfo{year}{2024}\natexlab{}.
\newblock \bibinfo{title}{Osiris: A Systolic Approach to Accelerating Fully Homomorphic Encryption}.
\newblock
\newblock
\showeprint[arxiv]{2408.09593}~[cs.CR]
\urldef\tempurl%
\url{https://arxiv.org/abs/2408.09593}
\showURL{%
\tempurl}


\bibitem[Feng et~al\mbox{.}(2024)]%
        {10.1145/3617232.3624852}
\bibfield{author}{\bibinfo{person}{Boyuan Feng}, \bibinfo{person}{Zheng Wang}, \bibinfo{person}{Yuke Wang}, \bibinfo{person}{Shu Yang}, {and} \bibinfo{person}{Yufei Ding}.} \bibinfo{year}{2024}\natexlab{}.
\newblock \showarticletitle{ZENO: A Type-based Optimization Framework for Zero Knowledge Neural Network Inference}. In \bibinfo{booktitle}{\emph{Proceedings of the 29th ACM International Conference on Architectural Support for Programming Languages and Operating Systems, Volume 1}} (La Jolla, CA, USA) \emph{(\bibinfo{series}{ASPLOS '24})}. \bibinfo{publisher}{Association for Computing Machinery}, \bibinfo{address}{New York, NY, USA}, \bibinfo{pages}{450–464}.
\newblock
\showISBNx{9798400703720}
\urldef\tempurl%
\url{https://doi.org/10.1145/3617232.3624852}
\showDOI{\tempurl}


\bibitem[Gabizon et~al\mbox{.}(2019)]%
        {plonk}
\bibfield{author}{\bibinfo{person}{Ariel Gabizon}, \bibinfo{person}{Zachary~J. Williamson}, {and} \bibinfo{person}{Oana Ciobotaru}.} \bibinfo{year}{2019}\natexlab{}.
\newblock \bibinfo{title}{{PLONK}: Permutations over Lagrange-bases for Oecumenical Noninteractive arguments of Knowledge}.
\newblock \bibinfo{howpublished}{Cryptology {ePrint} Archive, Paper 2019/953}.
\newblock
\urldef\tempurl%
\url{https://eprint.iacr.org/2019/953}
\showURL{%
\tempurl}


\bibitem[Garimella et~al\mbox{.}(2023)]%
        {karthik}
\bibfield{author}{\bibinfo{person}{Karthik Garimella}, \bibinfo{person}{Zahra Ghodsi}, \bibinfo{person}{Nandan~Kumar Jha}, \bibinfo{person}{Siddharth Garg}, {and} \bibinfo{person}{Brandon Reagen}.} \bibinfo{year}{2023}\natexlab{}.
\newblock \showarticletitle{Characterizing and Optimizing End-to-End Systems for Private Inference}. In \bibinfo{booktitle}{\emph{Proceedings of the 28th ACM International Conference on Architectural Support for Programming Languages and Operating Systems, Volume 3}} (Vancouver, BC, Canada) \emph{(\bibinfo{series}{ASPLOS 2023})}. \bibinfo{publisher}{Association for Computing Machinery}, \bibinfo{address}{New York, NY, USA}, \bibinfo{pages}{89–104}.
\newblock
\showISBNx{9781450399180}
\urldef\tempurl%
\url{https://doi.org/10.1145/3582016.3582065}
\showDOI{\tempurl}


\bibitem[Grassi et~al\mbox{.}(2019)]%
        {poseidon}
\bibfield{author}{\bibinfo{person}{Lorenzo Grassi}, \bibinfo{person}{Dmitry Khovratovich}, \bibinfo{person}{Christian Rechberger}, \bibinfo{person}{Arnab Roy}, {and} \bibinfo{person}{Markus Schofnegger}.} \bibinfo{year}{2019}\natexlab{}.
\newblock \bibinfo{title}{Poseidon: A New Hash Function for Zero-Knowledge Proof Systems}.
\newblock \bibinfo{howpublished}{Cryptology {ePrint} Archive, Paper 2019/458}.
\newblock
\urldef\tempurl%
\url{https://eprint.iacr.org/2019/458}
\showURL{%
\tempurl}


\bibitem[Groth(2016)]%
        {groth}
\bibfield{author}{\bibinfo{person}{Jens Groth}.} \bibinfo{year}{2016}\natexlab{}.
\newblock \bibinfo{title}{On the Size of Pairing-based Non-interactive Arguments}.
\newblock \bibinfo{howpublished}{Cryptology ePrint Archive, Paper 2016/260}.
\newblock
\urldef\tempurl%
\url{https://eprint.iacr.org/2016/260}
\showURL{%
\tempurl}
\newblock
\shownote{\url{https://eprint.iacr.org/2016/260}}.


\bibitem[Groth et~al\mbox{.}(2018)]%
        {universalsetup}
\bibfield{author}{\bibinfo{person}{Jens Groth}, \bibinfo{person}{Markulf Kohlweiss}, \bibinfo{person}{Mary Maller}, \bibinfo{person}{Sarah Meiklejohn}, {and} \bibinfo{person}{Ian Miers}.} \bibinfo{year}{2018}\natexlab{}.
\newblock \showarticletitle{Updatable and Universal Common Reference Strings with Applications to zk-SNARKs}. In \bibinfo{booktitle}{\emph{Advances in Cryptology - {CRYPTO} 2018 - 38th Annual International Cryptology Conference, Santa Barbara, CA, USA, August 19-23, 2018, Proceedings, Part {III}}} \emph{(\bibinfo{series}{Lecture Notes in Computer Science}, Vol.~\bibinfo{volume}{10993})}, \bibfield{editor}{\bibinfo{person}{Hovav Shacham} {and} \bibinfo{person}{Alexandra Boldyreva}} (Eds.). \bibinfo{publisher}{Springer}, \bibinfo{pages}{698--728}.
\newblock
\urldef\tempurl%
\url{https://doi.org/10.1007/978-3-319-96878-0\_24}
\showDOI{\tempurl}


\bibitem[Hirner et~al\mbox{.}(2025)]%
        {graz}
\bibfield{author}{\bibinfo{person}{Florian Hirner}, \bibinfo{person}{Florian Krieger}, {and} \bibinfo{person}{Sujoy~Sinha Roy}.} \bibinfo{year}{2025}\natexlab{}.
\newblock \bibinfo{title}{Chiplet-Based Techniques for Scalable and Memory-Aware Multi-Scalar Multiplication}.
\newblock \bibinfo{howpublished}{Cryptology {ePrint} Archive, Paper 2025/252}.
\newblock
\urldef\tempurl%
\url{https://eprint.iacr.org/2025/252}
\showURL{%
\tempurl}


\bibitem[Inc.(2020)]%
        {hbm2}
\bibfield{author}{\bibinfo{person}{Rambus Inc.}} \bibinfo{year}{2020}\natexlab{}.
\newblock \bibinfo{title}{White Paper: HBM2E and GDDR6: Memory Solutions for AI}.
\newblock \bibinfo{howpublished}{White Paper}.
\newblock


\bibitem[Jha and Reagen(2024)]%
        {jha2024deepreshape}
\bibfield{author}{\bibinfo{person}{Nandan~Kumar Jha} {and} \bibinfo{person}{Brandon Reagen}.} \bibinfo{year}{2024}\natexlab{}.
\newblock \showarticletitle{DeepReShape: Redesigning Neural Networks for Efficient Private Inference}.
\newblock \bibinfo{journal}{\emph{Transactions on Machine Learning Research (TMLR)}} (\bibinfo{year}{2024}).
\newblock


\bibitem[Ji et~al\mbox{.}(2024)]%
        {distMSM}
\bibfield{author}{\bibinfo{person}{Zhuoran Ji}, \bibinfo{person}{Zhiyuan Zhang}, \bibinfo{person}{Jiming Xu}, {and} \bibinfo{person}{Lei Ju}.} \bibinfo{year}{2024}\natexlab{}.
\newblock \showarticletitle{Accelerating Multi-Scalar Multiplication for Efficient Zero Knowledge Proofs with Multi-GPU Systems}. In \bibinfo{booktitle}{\emph{Proceedings of the 29th ACM International Conference on Architectural Support for Programming Languages and Operating Systems, Volume 3}} (La Jolla, CA, USA) \emph{(\bibinfo{series}{ASPLOS '24})}. \bibinfo{publisher}{Association for Computing Machinery}, \bibinfo{address}{New York, NY, USA}, \bibinfo{pages}{57–70}.
\newblock
\showISBNx{9798400703867}
\urldef\tempurl%
\url{https://doi.org/10.1145/3620666.3651364}
\showDOI{\tempurl}


\bibitem[Kim et~al\mbox{.}(2023)]%
        {sharp}
\bibfield{author}{\bibinfo{person}{Jongmin Kim}, \bibinfo{person}{Sangpyo Kim}, \bibinfo{person}{Jaewan Choi}, \bibinfo{person}{Jaiyoung Park}, \bibinfo{person}{Donghwan Kim}, {and} \bibinfo{person}{Jung~Ho Ahn}.} \bibinfo{year}{2023}\natexlab{}.
\newblock \showarticletitle{SHARP: A Short-Word Hierarchical Accelerator for Robust and Practical Fully Homomorphic Encryption}. In \bibinfo{booktitle}{\emph{Proceedings of the 50th Annual International Symposium on Computer Architecture}} (Orlando, FL, USA) \emph{(\bibinfo{series}{ISCA '23})}. \bibinfo{publisher}{Association for Computing Machinery}, \bibinfo{address}{New York, NY, USA}, Article \bibinfo{articleno}{18}, \bibinfo{numpages}{15}~pages.
\newblock
\showISBNx{9798400700958}
\urldef\tempurl%
\url{https://doi.org/10.1145/3579371.3589053}
\showDOI{\tempurl}


\bibitem[Kim et~al\mbox{.}(2022b)]%
        {ark}
\bibfield{author}{\bibinfo{person}{Jongmin Kim}, \bibinfo{person}{Gwangho Lee}, \bibinfo{person}{Sangpyo Kim}, \bibinfo{person}{Gina Sohn}, \bibinfo{person}{Minsoo Rhu}, \bibinfo{person}{John Kim}, {and} \bibinfo{person}{Jung~Ho Ahn}.} \bibinfo{year}{2022}\natexlab{b}.
\newblock \showarticletitle{ARK: Fully Homomorphic Encryption Accelerator with Runtime Data Generation and Inter-Operation Key Reuse}. In \bibinfo{booktitle}{\emph{2022 55th IEEE/ACM International Symposium on Microarchitecture (MICRO)}}. \bibinfo{pages}{1237--1254}.
\newblock
\urldef\tempurl%
\url{https://doi.org/10.1109/MICRO56248.2022.00086}
\showDOI{\tempurl}


\bibitem[Kim et~al\mbox{.}(2022a)]%
        {bts}
\bibfield{author}{\bibinfo{person}{Sangpyo Kim}, \bibinfo{person}{Jongmin Kim}, \bibinfo{person}{Michael~Jaemin Kim}, \bibinfo{person}{Wonkyung Jung}, \bibinfo{person}{John Kim}, \bibinfo{person}{Minsoo Rhu}, {and} \bibinfo{person}{Jung~Ho Ahn}.} \bibinfo{year}{2022}\natexlab{a}.
\newblock \showarticletitle{BTS: An Accelerator for Bootstrappable Fully Homomorphic Encryption}. In \bibinfo{booktitle}{\emph{Proceedings of the 49th Annual International Symposium on Computer Architecture}} (New York, New York) \emph{(\bibinfo{series}{ISCA '22})}. \bibinfo{publisher}{Association for Computing Machinery}, \bibinfo{address}{New York, NY, USA}, \bibinfo{pages}{711–725}.
\newblock
\showISBNx{9781450386104}
\urldef\tempurl%
\url{https://doi.org/10.1145/3470496.3527415}
\showDOI{\tempurl}


\bibitem[Kohrita and Towa(2024)]%
        {zeromorph}
\bibfield{author}{\bibinfo{person}{Tohru Kohrita} {and} \bibinfo{person}{Patrick Towa}.} \bibinfo{year}{2024}\natexlab{}.
\newblock \showarticletitle{Zeromorph: Zero-Knowledge Multilinear-Evaluation Proofs from Homomorphic Univariate Commitments}.
\newblock \bibinfo{journal}{\emph{J. Cryptol.}} \bibinfo{volume}{37}, \bibinfo{number}{4} (\bibinfo{year}{2024}), \bibinfo{pages}{38}.
\newblock
\urldef\tempurl%
\url{https://doi.org/10.1007/S00145-024-09519-0}
\showDOI{\tempurl}


\bibitem[Kothapalli et~al\mbox{.}(2022)]%
        {nova}
\bibfield{author}{\bibinfo{person}{Abhiram Kothapalli}, \bibinfo{person}{Srinath T.~V. Setty}, {and} \bibinfo{person}{Ioanna Tzialla}.} \bibinfo{year}{2022}\natexlab{}.
\newblock \showarticletitle{Nova: Recursive Zero-Knowledge Arguments from Folding Schemes}. In \bibinfo{booktitle}{\emph{Advances in Cryptology - {CRYPTO} 2022 - 42nd Annual International Cryptology Conference, {CRYPTO} 2022, Santa Barbara, CA, USA, August 15-18, 2022, Proceedings, Part {IV}}} \emph{(\bibinfo{series}{Lecture Notes in Computer Science}, Vol.~\bibinfo{volume}{13510})}, \bibfield{editor}{\bibinfo{person}{Yevgeniy Dodis} {and} \bibinfo{person}{Thomas Shrimpton}} (Eds.). \bibinfo{publisher}{Springer}, \bibinfo{pages}{359--388}.
\newblock
\urldef\tempurl%
\url{https://doi.org/10.1007/978-3-031-15985-5\_13}
\showDOI{\tempurl}


\bibitem[Liu et~al\mbox{.}(2024a)]%
        {priorMSM}
\bibfield{author}{\bibinfo{person}{Changxu Liu}, \bibinfo{person}{Hao Zhou}, \bibinfo{person}{Patrick Dai}, \bibinfo{person}{Li Shang}, {and} \bibinfo{person}{Fan Yang}.} \bibinfo{year}{2024}\natexlab{a}.
\newblock \showarticletitle{PriorMSM: An Efficient Acceleration Architecture for Multi-Scalar Multiplication}.
\newblock \bibinfo{journal}{\emph{ACM Transactions on Design Automation of Electronic Systems}} \bibinfo{volume}{29}, \bibinfo{number}{5} (\bibinfo{year}{2024}), \bibinfo{pages}{1--26}.
\newblock


\bibitem[Liu et~al\mbox{.}(2024b)]%
        {myotosis}
\bibfield{author}{\bibinfo{person}{Changxu Liu}, \bibinfo{person}{Hao Zhou}, \bibinfo{person}{Lan Yang}, \bibinfo{person}{Zheng Wu}, \bibinfo{person}{Patrick Dai}, \bibinfo{person}{Yinlong Li}, \bibinfo{person}{Shiyong Wu}, {and} \bibinfo{person}{Fan Yang}.} \bibinfo{year}{2024}\natexlab{b}.
\newblock \showarticletitle{Myosotis: An Efficiently Pipelined and Parameterized Multi-Scalar Multiplication Architecture via Data Sharing}.
\newblock \bibinfo{journal}{\emph{IEEE Transactions on Computer-Aided Design of Integrated Circuits and Systems}} (\bibinfo{year}{2024}), \bibinfo{pages}{1--1}.
\newblock
\urldef\tempurl%
\url{https://doi.org/10.1109/TCAD.2024.3524364}
\showDOI{\tempurl}


\bibitem[Liu et~al\mbox{.}(2024c)]%
        {gypso}
\bibfield{author}{\bibinfo{person}{Changxu Liu}, \bibinfo{person}{Hao Zhou}, \bibinfo{person}{Lan Yang}, \bibinfo{person}{Jiamin Xu}, \bibinfo{person}{Patrick Dai}, {and} \bibinfo{person}{Fan Yang}.} \bibinfo{year}{2024}\natexlab{c}.
\newblock \showarticletitle{Gypsophila: A Scalable and Bandwidth-Optimized Multi-Scalar Multiplication Architecture}. In \bibinfo{booktitle}{\emph{Proceedings of the 61st ACM/IEEE Design Automation Conference}} (San Francisco, CA, USA) \emph{(\bibinfo{series}{DAC '24})}. \bibinfo{publisher}{Association for Computing Machinery}, \bibinfo{address}{New York, NY, USA}, Article \bibinfo{articleno}{94}, \bibinfo{numpages}{6}~pages.
\newblock
\showISBNx{9798400706011}
\urldef\tempurl%
\url{https://doi.org/10.1145/3649329.3658259}
\showDOI{\tempurl}


\bibitem[Lu et~al\mbox{.}(2024)]%
        {batchzk}
\bibfield{author}{\bibinfo{person}{Tao Lu}, \bibinfo{person}{Yuxun Chen}, \bibinfo{person}{Zonghui Wang}, \bibinfo{person}{Xiaohang Wang}, \bibinfo{person}{Wenzhi Chen}, {and} \bibinfo{person}{Jiaheng Zhang}.} \bibinfo{year}{2024}\natexlab{}.
\newblock \bibinfo{title}{{BatchZK}: A Fully Pipelined {GPU}-Accelerated System for Batch Generation of Zero-Knowledge Proofs}.
\newblock \bibinfo{howpublished}{Cryptology {ePrint} Archive, Paper 2024/1862}.
\newblock
\urldef\tempurl%
\url{https://eprint.iacr.org/2024/1862}
\showURL{%
\tempurl}


\bibitem[Lu et~al\mbox{.}(2022)]%
        {cuZK}
\bibfield{author}{\bibinfo{person}{Tao Lu}, \bibinfo{person}{Chengkun Wei}, \bibinfo{person}{Ruijing Yu}, \bibinfo{person}{Chaochao Chen}, \bibinfo{person}{Wenjing Fang}, \bibinfo{person}{Lei Wang}, \bibinfo{person}{Zeke Wang}, {and} \bibinfo{person}{Wenzhi Chen}.} \bibinfo{year}{2022}\natexlab{}.
\newblock \bibinfo{title}{cuZK: Accelerating Zero-Knowledge Proof with A Faster Parallel Multi-Scalar Multiplication Algorithm on GPUs}.
\newblock \bibinfo{howpublished}{Cryptology ePrint Archive, Paper 2022/1321}.
\newblock
\urldef\tempurl%
\url{https://eprint.iacr.org/2022/1321}
\showURL{%
\tempurl}
\newblock
\shownote{\url{https://eprint.iacr.org/2022/1321}}.


\bibitem[Ma et~al\mbox{.}(2023)]%
        {gzkp}
\bibfield{author}{\bibinfo{person}{Weiliang Ma}, \bibinfo{person}{Qian Xiong}, \bibinfo{person}{Xuanhua Shi}, \bibinfo{person}{Xiaosong Ma}, \bibinfo{person}{Hai Jin}, \bibinfo{person}{Haozhao Kuang}, \bibinfo{person}{Mingyu Gao}, \bibinfo{person}{Ye Zhang}, \bibinfo{person}{Haichen Shen}, {and} \bibinfo{person}{Weifang Hu}.} \bibinfo{year}{2023}\natexlab{}.
\newblock \showarticletitle{GZKP: A GPU Accelerated Zero-Knowledge Proof System}. In \bibinfo{booktitle}{\emph{Proceedings of the 28th ACM International Conference on Architectural Support for Programming Languages and Operating Systems, Volume 2}} (Vancouver, BC, Canada) \emph{(\bibinfo{series}{ASPLOS 2023})}. \bibinfo{publisher}{Association for Computing Machinery}, \bibinfo{address}{New York, NY, USA}, \bibinfo{pages}{340–353}.
\newblock
\showISBNx{9781450399166}
\urldef\tempurl%
\url{https://doi.org/10.1145/3575693.3575711}
\showDOI{\tempurl}


\bibitem[Mo et~al\mbox{.}(2025a)]%
        {mulfunc_tree}
\bibfield{author}{\bibinfo{person}{Jianqiao Mo}, \bibinfo{person}{Alhad Daftardar}, \bibinfo{person}{Joey Ah-kiow}, \bibinfo{person}{Kaiyue Guo}, \bibinfo{person}{Benedikt Bünz}, \bibinfo{person}{Siddharth Garg}, {and} \bibinfo{person}{Brandon Reagen}.} \bibinfo{year}{2025}\natexlab{a}.
\newblock \bibinfo{title}{MTU: The Multifunction Tree Unit in zkSpeed for Accelerating HyperPlonk}.
\newblock
\newblock
\showeprint[arxiv]{2507.16793}~[cs.AR]
\urldef\tempurl%
\url{https://arxiv.org/abs/2507.16793}
\showURL{%
\tempurl}


\bibitem[Mo et~al\mbox{.}(2023)]%
        {haac}
\bibfield{author}{\bibinfo{person}{Jianqiao Mo}, \bibinfo{person}{Jayanth Gopinath}, {and} \bibinfo{person}{Brandon Reagen}.} \bibinfo{year}{2023}\natexlab{}.
\newblock \showarticletitle{HAAC: A Hardware-Software Co-Design to Accelerate Garbled Circuits}. In \bibinfo{booktitle}{\emph{Proceedings of the 50th Annual International Symposium on Computer Architecture}} (Orlando, FL, USA) \emph{(\bibinfo{series}{ISCA '23})}. \bibinfo{publisher}{Association for Computing Machinery}, \bibinfo{address}{New York, NY, USA}, Article \bibinfo{articleno}{10}, \bibinfo{numpages}{13}~pages.
\newblock
\showISBNx{9798400700958}
\urldef\tempurl%
\url{https://doi.org/10.1145/3579371.3589045}
\showDOI{\tempurl}


\bibitem[Mo et~al\mbox{.}(2025b)]%
        {mo2025able}
\bibfield{author}{\bibinfo{person}{Jianqiao~Cambridge Mo}, \bibinfo{person}{Karthik Garimella}, \bibinfo{person}{Austin Ebel}, {and} \bibinfo{person}{Brandon Reagen}.} \bibinfo{year}{2025}\natexlab{b}.
\newblock \bibinfo{title}{{ABLE}: Optimizing Mixed Arithmetic and Boolean Garbled Circuit}.
\newblock \bibinfo{howpublished}{Cryptology {ePrint} Archive, Paper 2025/048}.
\newblock
\urldef\tempurl%
\url{https://eprint.iacr.org/2025/048}
\showURL{%
\tempurl}


\bibitem[Monnot(2024)]%
        {ethereum_blocksize_2024}
\bibfield{author}{\bibinfo{person}{Barnabé Monnot}.} \bibinfo{year}{2024}\natexlab{}.
\newblock \bibinfo{title}{On Block Sizes, Gas Limits, and Scalability}.
\newblock \bibinfo{howpublished}{\url{https://ethresear.ch/t/on-block-sizes-gas-limits-and-scalability/18444}}.
\newblock
\newblock
\shownote{Accessed: 2025-05-06}.


\bibitem[Montgomery(1987)]%
        {Montgomery_1987}
\bibfield{author}{\bibinfo{person}{Peter~L. Montgomery}.} \bibinfo{year}{1987}\natexlab{}.
\newblock \showarticletitle{Speeding the Pollard and elliptic curve methods of factorization}.
\newblock \bibinfo{journal}{\emph{Math. Comp.}} \bibinfo{volume}{48}, \bibinfo{number}{177} (\bibinfo{year}{1987}), \bibinfo{pages}{243–264}.
\newblock
\showISSN{0025-5718, 1088-6842}
\urldef\tempurl%
\url{https://doi.org/10.1090/S0025-5718-1987-0866113-7}
\showDOI{\tempurl}


\bibitem[Morgan(2019a)]%
        {cpu_amd1}
\bibfield{author}{\bibinfo{person}{Timothy~Prickett Morgan}.} \bibinfo{year}{2019}\natexlab{a}.
\newblock \bibinfo{title}{AMD Doubles Down – And Up – With Rome Epyc Server Chips}.
\newblock
\newblock
\urldef\tempurl%
\url{https://www.nextplatform.com/2019/08/07/amd-doubles-down-and-up-with-rome-epyc-server-chips/}
\showURL{%
\tempurl}


\bibitem[Morgan(2019b)]%
        {amd_article}
\bibfield{author}{\bibinfo{person}{Timothy~Prickett Morgan}.} \bibinfo{year}{2019}\natexlab{b}.
\newblock \bibinfo{booktitle}{\emph{A Deep Dive Into AMD's Rome EPYC Architecture}}.
\newblock
\urldef\tempurl%
\url{https://www.nextplatform.com/2019/08/15/a-deep-dive-into-amds-rome-epyc-architecture/}
\showURL{%
\tempurl}


\bibitem[Neda et~al\mbox{.}(2024)]%
        {ciflow}
\bibfield{author}{\bibinfo{person}{Negar Neda}, \bibinfo{person}{Austin Ebel}, \bibinfo{person}{Benedict Reynwar}, {and} \bibinfo{person}{Brandon Reagen}.} \bibinfo{year}{2024}\natexlab{}.
\newblock \showarticletitle{CiFlow: Dataflow Analysis and Optimization of Key Switching for Homomorphic Encryption}. In \bibinfo{booktitle}{\emph{2024 IEEE International Symposium on Performance Analysis of Systems and Software (ISPASS)}}. \bibinfo{pages}{61--72}.
\newblock
\urldef\tempurl%
\url{https://doi.org/10.1109/ISPASS61541.2024.00016}
\showDOI{\tempurl}


\bibitem[Nelson(2022)]%
        {trusted_set_up}
\bibfield{author}{\bibinfo{person}{Jason Nelson}.} \bibinfo{year}{2022}\natexlab{}.
\newblock \bibinfo{title}{Zcash Nixes Trusted Setup, Enters New Era With Major Network Update}.
\newblock
\newblock
\urldef\tempurl%
\url{https://decrypt.co/101762/zcash-nixes-trusted-setup-enters-new-era-with-major-network-update}
\showURL{%
\tempurl}


\bibitem[{OpenCores}(2013)]%
        {opencores_sha3}
\bibfield{author}{\bibinfo{person}{{OpenCores}}.} \bibinfo{year}{2013}\natexlab{}.
\newblock \bibinfo{title}{SHA-3 IP Core}.
\newblock \bibinfo{howpublished}{\url{https://opencores.org/projects/sha3}}.
\newblock
\newblock
\shownote{Accessed: November 16, 2024}.


\bibitem[Pippenger(1976)]%
        {pippenger}
\bibfield{author}{\bibinfo{person}{Nicholas Pippenger}.} \bibinfo{year}{1976}\natexlab{}.
\newblock \showarticletitle{On the evaluation of powers and related problems}. In \bibinfo{booktitle}{\emph{17th Annual Symposium on Foundations of Computer Science (sfcs 1976)}}. \bibinfo{pages}{258--263}.
\newblock
\urldef\tempurl%
\url{https://doi.org/10.1109/SFCS.1976.21}
\showDOI{\tempurl}


\bibitem[Pornin(2020)]%
        {Pornin_2020}
\bibfield{author}{\bibinfo{person}{Thomas Pornin}.} \bibinfo{year}{2020}\natexlab{}.
\newblock \showarticletitle{Optimized Binary GCD for Modular Inversion}.
\newblock  \bibinfo{number}{2020/972} (\bibinfo{year}{2020}).
\newblock
\urldef\tempurl%
\url{https://eprint.iacr.org/2020/972}
\showURL{%
\tempurl}


\bibitem[Qiu et~al\mbox{.}(2024)]%
        {MSMAC}
\bibfield{author}{\bibinfo{person}{Pengcheng Qiu}, \bibinfo{person}{Guiming Wu}, \bibinfo{person}{Tingqiang Chu}, \bibinfo{person}{Changzheng Wei}, \bibinfo{person}{Runzhou Luo}, \bibinfo{person}{Ying Yan}, \bibinfo{person}{Wei Wang}, {and} \bibinfo{person}{Hui Zhang}.} \bibinfo{year}{2024}\natexlab{}.
\newblock \showarticletitle{MSMAC: Accelerating Multi-Scalar Multiplication for Zero-Knowledge Proof}. In \bibinfo{booktitle}{\emph{Proceedings of the 61st ACM/IEEE Design Automation Conference}} (San Francisco, CA, USA) \emph{(\bibinfo{series}{DAC '24})}. \bibinfo{publisher}{Association for Computing Machinery}, \bibinfo{address}{New York, NY, USA}, Article \bibinfo{articleno}{66}, \bibinfo{numpages}{6}~pages.
\newblock
\showISBNx{9798400706011}
\urldef\tempurl%
\url{https://doi.org/10.1145/3649329.3655672}
\showDOI{\tempurl}


\bibitem[Samardzic et~al\mbox{.}(2021)]%
        {f1}
\bibfield{author}{\bibinfo{person}{Nikola Samardzic}, \bibinfo{person}{Axel Feldmann}, \bibinfo{person}{Aleksandar Krastev}, \bibinfo{person}{Srinivas Devadas}, \bibinfo{person}{Ronald Dreslinski}, \bibinfo{person}{Christopher Peikert}, {and} \bibinfo{person}{Daniel Sanchez}.} \bibinfo{year}{2021}\natexlab{}.
\newblock \showarticletitle{F1: A Fast and Programmable Accelerator for Fully Homomorphic Encryption}. In \bibinfo{booktitle}{\emph{MICRO-54: 54th Annual IEEE/ACM International Symposium on Microarchitecture}} (Virtual Event, Greece) \emph{(\bibinfo{series}{MICRO '21})}. \bibinfo{publisher}{Association for Computing Machinery}, \bibinfo{address}{New York, NY, USA}, \bibinfo{pages}{238–252}.
\newblock
\showISBNx{9781450385572}
\urldef\tempurl%
\url{https://doi.org/10.1145/3466752.3480070}
\showDOI{\tempurl}


\bibitem[Samardzic et~al\mbox{.}(2022)]%
        {clake}
\bibfield{author}{\bibinfo{person}{Nikola Samardzic}, \bibinfo{person}{Axel Feldmann}, \bibinfo{person}{Aleksandar Krastev}, \bibinfo{person}{Nathan Manohar}, \bibinfo{person}{Nicholas Genise}, \bibinfo{person}{Srinivas Devadas}, \bibinfo{person}{Karim Eldefrawy}, \bibinfo{person}{Chris Peikert}, {and} \bibinfo{person}{Daniel Sanchez}.} \bibinfo{year}{2022}\natexlab{}.
\newblock \showarticletitle{CraterLake: A Hardware Accelerator for Efficient Unbounded Computation on Encrypted Data}. In \bibinfo{booktitle}{\emph{Proceedings of the 49th Annual International Symposium on Computer Architecture}} (New York, New York) \emph{(\bibinfo{series}{ISCA '22})}. \bibinfo{publisher}{Association for Computing Machinery}, \bibinfo{address}{New York, NY, USA}, \bibinfo{pages}{173–187}.
\newblock
\showISBNx{9781450386104}
\urldef\tempurl%
\url{https://doi.org/10.1145/3470496.3527393}
\showDOI{\tempurl}


\bibitem[Samardzic et~al\mbox{.}(2024)]%
        {nocap}
\bibfield{author}{\bibinfo{person}{Nikola Samardzic}, \bibinfo{person}{Simon Langowski}, \bibinfo{person}{Srinivas Devadas}, {and} \bibinfo{person}{Daniel Sanchez}.} \bibinfo{year}{2024}\natexlab{}.
\newblock \bibinfo{title}{Accelerating Zero-Knowledge Proofs Through Hardware-Algorithm Co-Design}.
\newblock \bibinfo{howpublished}{Preprint}.
\newblock
\newblock
\shownote{https://people.csail.mit.edu/devadas/pubs/micro24\_nocap.pdf}.


\bibitem[Schlachter and Drake(2019)]%
        {micron_ddr5}
\bibfield{author}{\bibinfo{person}{Scott Schlachter} {and} \bibinfo{person}{Brian Drake}.} \bibinfo{year}{2019}\natexlab{}.
\newblock \showarticletitle{Introducing Micron{\textregistered} DDR5 SDRAM: More than a generational update}.
\newblock \bibinfo{journal}{\emph{XP055844818}}  \bibinfo{volume}{31} (\bibinfo{year}{2019}), \bibinfo{pages}{6}.
\newblock


\bibitem[Setty(2020)]%
        {spartan}
\bibfield{author}{\bibinfo{person}{Srinath Setty}.} \bibinfo{year}{2020}\natexlab{}.
\newblock \showarticletitle{Spartan: Efficient and General-Purpose zkSNARKs Without Trusted Setup}. In \bibinfo{booktitle}{\emph{Advances in Cryptology -- CRYPTO 2020}}, \bibfield{editor}{\bibinfo{person}{Daniele Micciancio} {and} \bibinfo{person}{Thomas Ristenpart}} (Eds.). \bibinfo{publisher}{Springer International Publishing}, \bibinfo{address}{Cham}, \bibinfo{pages}{704--737}.
\newblock
\showISBNx{978-3-030-56877-1}


\bibitem[Setty et~al\mbox{.}(2024)]%
        {lasso}
\bibfield{author}{\bibinfo{person}{Srinath T.~V. Setty}, \bibinfo{person}{Justin Thaler}, {and} \bibinfo{person}{Riad~S. Wahby}.} \bibinfo{year}{2024}\natexlab{}.
\newblock \showarticletitle{Unlocking the Lookup Singularity with Lasso}. In \bibinfo{booktitle}{\emph{Advances in Cryptology - {EUROCRYPT} 2024 - 43rd Annual International Conference on the Theory and Applications of Cryptographic Techniques, Zurich, Switzerland, May 26-30, 2024, Proceedings, Part {VI}}} \emph{(\bibinfo{series}{Lecture Notes in Computer Science}, Vol.~\bibinfo{volume}{14656})}, \bibfield{editor}{\bibinfo{person}{Marc Joye} {and} \bibinfo{person}{Gregor Leander}} (Eds.). \bibinfo{publisher}{Springer}, \bibinfo{pages}{180--209}.
\newblock
\urldef\tempurl%
\url{https://doi.org/10.1007/978-3-031-58751-1\_7}
\showDOI{\tempurl}


\bibitem[Sheybani et~al\mbox{.}(2025)]%
        {gottahashemall}
\bibfield{author}{\bibinfo{person}{Nojan Sheybani}, \bibinfo{person}{Tengkai Gong}, \bibinfo{person}{Anees Ahmed}, \bibinfo{person}{Nges~Brian Njungle}, \bibinfo{person}{Michel Kinsy}, {and} \bibinfo{person}{Farinaz Koushanfar}.} \bibinfo{year}{2025}\natexlab{}.
\newblock \bibinfo{title}{Gotta Hash 'Em All! Speeding Up Hash Functions for Zero-Knowledge Proof Applications}.
\newblock
\newblock
\showeprint[arxiv]{2501.18780}~[cs.CR]
\urldef\tempurl%
\url{https://arxiv.org/abs/2501.18780}
\showURL{%
\tempurl}


\bibitem[Soni et~al\mbox{.}(2023)]%
        {rpu}
\bibfield{author}{\bibinfo{person}{Deepraj Soni}, \bibinfo{person}{Negar Neda}, \bibinfo{person}{Naifeng Zhang}, \bibinfo{person}{Benedict Reynwar}, \bibinfo{person}{Homer Gamil}, \bibinfo{person}{Benjamin Heyman}, \bibinfo{person}{Mohammed Nabeel}, \bibinfo{person}{Ahmad~Al Badawi}, \bibinfo{person}{Yuriy Polyakov}, \bibinfo{person}{Kellie Canida}, \bibinfo{person}{Massoud Pedram}, \bibinfo{person}{Michail Maniatakos}, \bibinfo{person}{David~Bruce Cousins}, \bibinfo{person}{Franz Franchetti}, \bibinfo{person}{Matthew French}, \bibinfo{person}{Andrew Schmidt}, {and} \bibinfo{person}{Brandon Reagen}.} \bibinfo{year}{2023}\natexlab{}.
\newblock \showarticletitle{RPU: The Ring Processing Unit}. In \bibinfo{booktitle}{\emph{2023 IEEE International Symposium on Performance Analysis of Systems and Software (ISPASS)}}. \bibinfo{pages}{272--282}.
\newblock
\urldef\tempurl%
\url{https://doi.org/10.1109/ISPASS57527.2023.00034}
\showDOI{\tempurl}


\bibitem[tech(2024)]%
        {cpu_amd3}
\bibfield{author}{\bibinfo{person}{powerup tech}.} \bibinfo{year}{2024}\natexlab{}.
\newblock \bibinfo{title}{AMD EPYC 7502 Specs}.
\newblock
\newblock
\urldef\tempurl%
\url{https://www.techpowerup.com/cpu-specs/epyc-7502.c2250}
\showURL{%
\tempurl}


\bibitem[Thaler(2013)]%
        {gkr_paper}
\bibfield{author}{\bibinfo{person}{Justin Thaler}.} \bibinfo{year}{2013}\natexlab{}.
\newblock \bibinfo{title}{Time-Optimal Interactive Proofs for Circuit Evaluation}.
\newblock \bibinfo{howpublished}{Cryptology {ePrint} Archive, Paper 2013/351}.
\newblock
\urldef\tempurl%
\url{https://eprint.iacr.org/2013/351}
\showURL{%
\tempurl}


\bibitem[Thaler(2022)]%
        {thaler_proofs_args_zk}
\bibfield{author}{\bibinfo{person}{Justin Thaler}.} \bibinfo{year}{2022}\natexlab{}.
\newblock \bibinfo{booktitle}{\emph{Proofs, Arguments, and Zero-Knowledge}}.
\newblock
\urldef\tempurl%
\url{https://people.cs.georgetown.edu/jthaler/ProofsArgsAndZK.pdf}
\showURL{%
\tempurl}


\bibitem[Valiant(2008)]%
        {ivc}
\bibfield{author}{\bibinfo{person}{Paul Valiant}.} \bibinfo{year}{2008}\natexlab{}.
\newblock \showarticletitle{Incrementally Verifiable Computation or Proofs of Knowledge Imply Time/Space Efficiency}. In \bibinfo{booktitle}{\emph{Theory of Cryptography, Fifth Theory of Cryptography Conference, {TCC} 2008, New York, USA, March 19-21, 2008}} \emph{(\bibinfo{series}{Lecture Notes in Computer Science}, Vol.~\bibinfo{volume}{4948})}, \bibfield{editor}{\bibinfo{person}{Ran Canetti}} (Ed.). \bibinfo{publisher}{Springer}, \bibinfo{pages}{1--18}.
\newblock
\urldef\tempurl%
\url{https://doi.org/10.1007/978-3-540-78524-8\_1}
\showDOI{\tempurl}


\bibitem[Wang and Gao(2023)]%
        {sam}
\bibfield{author}{\bibinfo{person}{Cheng Wang} {and} \bibinfo{person}{Mingyu Gao}.} \bibinfo{year}{2023}\natexlab{}.
\newblock \showarticletitle{SAM: A Scalable Accelerator for Number Theoretic Transform Using Multi-Dimensional Decomposition}. In \bibinfo{booktitle}{\emph{2023 IEEE/ACM International Conference on Computer Aided Design (ICCAD)}}. \bibinfo{pages}{1--9}.
\newblock
\urldef\tempurl%
\url{https://doi.org/10.1109/ICCAD57390.2023.10323744}
\showDOI{\tempurl}


\bibitem[Wang and Gao(2025)]%
        {unizk}
\bibfield{author}{\bibinfo{person}{Cheng Wang} {and} \bibinfo{person}{Mingyu Gao}.} \bibinfo{year}{2025}\natexlab{}.
\newblock \showarticletitle{UniZK: Accelerating Zero-Knowledge Proof with Unified Hardware and Flexible Kernel Mapping}. In \bibinfo{booktitle}{\emph{Proceedings of the 30th ACM International Conference on Architectural Support for Programming Languages and Operating Systems, Volume 1}} (Rotterdam, Netherlands) \emph{(\bibinfo{series}{ASPLOS '25})}. \bibinfo{publisher}{Association for Computing Machinery}, \bibinfo{address}{New York, NY, USA}, \bibinfo{pages}{1101–1117}.
\newblock
\showISBNx{9798400706981}
\urldef\tempurl%
\url{https://doi.org/10.1145/3669940.3707228}
\showDOI{\tempurl}


\bibitem[Xie et~al\mbox{.}(2022)]%
        {orion}
\bibfield{author}{\bibinfo{person}{Tiancheng Xie}, \bibinfo{person}{Yupeng Zhang}, {and} \bibinfo{person}{Dawn Song}.} \bibinfo{year}{2022}\natexlab{}.
\newblock \showarticletitle{Orion: Zero Knowledge Proof with Linear Prover Time}. In \bibinfo{booktitle}{\emph{Advances in Cryptology - {CRYPTO} 2022 - 42nd Annual International Cryptology Conference, {CRYPTO} 2022, Santa Barbara, CA, USA, August 15-18, 2022, Proceedings, Part {IV}}} \emph{(\bibinfo{series}{Lecture Notes in Computer Science}, Vol.~\bibinfo{volume}{13510})}, \bibfield{editor}{\bibinfo{person}{Yevgeniy Dodis} {and} \bibinfo{person}{Thomas Shrimpton}} (Eds.). \bibinfo{publisher}{Springer}, \bibinfo{pages}{299--328}.
\newblock
\urldef\tempurl%
\url{https://doi.org/10.1007/978-3-031-15985-5\_11}
\showDOI{\tempurl}


\bibitem[Xiong et~al\mbox{.}(2023)]%
        {verizexe}
\bibfield{author}{\bibinfo{person}{Alex~Luoyuan Xiong}, \bibinfo{person}{Binyi Chen}, \bibinfo{person}{Zhenfei Zhang}, \bibinfo{person}{Benedikt B{\"{u}}nz}, \bibinfo{person}{Ben Fisch}, \bibinfo{person}{Fernando Krell}, {and} \bibinfo{person}{Philippe Camacho}.} \bibinfo{year}{2023}\natexlab{}.
\newblock \showarticletitle{VeriZexe: Decentralized Private Computation with Universal Setup}. In \bibinfo{booktitle}{\emph{32nd {USENIX} Security Symposium, {USENIX} Security 2023, Anaheim, CA, USA, August 9-11, 2023}}, \bibfield{editor}{\bibinfo{person}{Joseph~A. Calandrino} {and} \bibinfo{person}{Carmela Troncoso}} (Eds.). \bibinfo{publisher}{{USENIX} Association}, \bibinfo{pages}{4445--4462}.
\newblock
\urldef\tempurl%
\url{https://www.usenix.org/conference/usenixsecurity23/presentation/xiong}
\showURL{%
\tempurl}


\bibitem[Yang et~al\mbox{.}(2025)]%
        {legozk}
\bibfield{author}{\bibinfo{person}{Zhengbang Yang}, \bibinfo{person}{Lutan Zhao}, \bibinfo{person}{Peinan Li}, \bibinfo{person}{Han Liu}, \bibinfo{person}{Kai Li}, \bibinfo{person}{Boyan Zhao}, \bibinfo{person}{Dan Meng}, {and} \bibinfo{person}{Rui Hou}.} \bibinfo{year}{2025}\natexlab{}.
\newblock \showarticletitle{LegoZK: A Dynamically Reconfigurable Accelerator for Zero-Knowledge Proof}. In \bibinfo{booktitle}{\emph{2025 IEEE International Symposium on High Performance Computer Architecture (HPCA)}}. \bibinfo{pages}{113--126}.
\newblock
\urldef\tempurl%
\url{https://doi.org/10.1109/HPCA61900.2025.00020}
\showDOI{\tempurl}


\bibitem[Zhang and Franchetti(2025)]%
        {zhang2025code}
\bibfield{author}{\bibinfo{person}{Naifeng Zhang} {and} \bibinfo{person}{Franz Franchetti}.} \bibinfo{year}{2025}\natexlab{}.
\newblock \showarticletitle{Code Generation for Cryptographic Kernels using Multi-word Modular Arithmetic on GPU}. In \bibinfo{booktitle}{\emph{Proceedings of the 23rd ACM/IEEE International Symposium on Code Generation and Optimization}} (Las Vegas, NV, USA) \emph{(\bibinfo{series}{CGO '25})}. \bibinfo{publisher}{Association for Computing Machinery}, \bibinfo{address}{New York, NY, USA}, \bibinfo{pages}{476–492}.
\newblock
\showISBNx{9798400712753}
\urldef\tempurl%
\url{https://doi.org/10.1145/3696443.3708948}
\showDOI{\tempurl}


\bibitem[Zhang et~al\mbox{.}(2021)]%
        {pipezk}
\bibfield{author}{\bibinfo{person}{Ye Zhang}, \bibinfo{person}{Shuo Wang}, \bibinfo{person}{Xian Zhang}, \bibinfo{person}{Jiangbin Dong}, \bibinfo{person}{Xingzhong Mao}, \bibinfo{person}{Fan Long}, \bibinfo{person}{Cong Wang}, \bibinfo{person}{Dong Zhou}, \bibinfo{person}{Mingyu Gao}, {and} \bibinfo{person}{Guangyu Sun}.} \bibinfo{year}{2021}\natexlab{}.
\newblock \showarticletitle{PipeZK: Accelerating Zero-Knowledge Proof with a Pipelined Architecture}. In \bibinfo{booktitle}{\emph{2021 ACM/IEEE 48th Annual International Symposium on Computer Architecture (ISCA)}}.
\newblock


\bibitem[Zhang et~al\mbox{.}(2022)]%
        {hyperplonkespresso}
\bibfield{author}{\bibinfo{person}{Zhenfei Zhang}, \bibinfo{person}{Binyi Chen}, \bibinfo{person}{Benedikt B\"unz}, {and} \bibinfo{person}{Alex Xiong}.} \bibinfo{year}{2022}\natexlab{}.
\newblock \bibinfo{title}{Hyperplonk Implementation}.
\newblock
\newblock
\urldef\tempurl%
\url{https://github.com/EspressoSystems/hyperplonk}
\showURL{%
\tempurl}


\bibitem[Zhou et~al\mbox{.}(2024)]%
        {reZK}
\bibfield{author}{\bibinfo{person}{Hao Zhou}, \bibinfo{person}{Changxu Liu}, \bibinfo{person}{Lan Yang}, \bibinfo{person}{Li Shang}, {and} \bibinfo{person}{Fan Yang}.} \bibinfo{year}{2024}\natexlab{}.
\newblock \showarticletitle{ReZK: A Highly Reconfigurable Accelerator for Zero-Knowledge Proof}.
\newblock \bibinfo{journal}{\emph{IEEE Transactions on Circuits and Systems I: Regular Papers}} (\bibinfo{year}{2024}).
\newblock


\bibitem[Zhu et~al\mbox{.}(2024)]%
        {elastic_msm}
\bibfield{author}{\bibinfo{person}{Xudong Zhu}, \bibinfo{person}{Haoqi He}, \bibinfo{person}{Zhengbang Yang}, \bibinfo{person}{Yi Deng}, \bibinfo{person}{Lutan Zhao}, {and} \bibinfo{person}{Rui Hou}.} \bibinfo{year}{2024}\natexlab{}.
\newblock \bibinfo{title}{Elastic {MSM}: A Fast, Elastic and Modular Preprocessing Technique for Multi-Scalar Multiplication Algorithm on {GPUs}}.
\newblock \bibinfo{howpublished}{Cryptology {ePrint} Archive, Paper 2024/057}.
\newblock
\urldef\tempurl%
\url{https://eprint.iacr.org/2024/057}
\showURL{%
\tempurl}


\end{thebibliography}


% \appendix

\end{document}